\documentclass[a4paper,accepted=2026-03-09]{quantumarticle}
\pdfoutput=1 

\usepackage{color,amsthm,amsmath,amsxtra,amsfonts,dsfont,graphicx,bm,amssymb}
\usepackage[colorlinks=true,linkcolor=blue, citecolor=blue, urlcolor=blue, bookmarks]{hyperref}
\usepackage{centernot}
\usepackage[dvipsnames]{xcolor}
\usepackage{graphicx}
\usepackage{tikz}
\usepackage{braket}
\usepackage{multirow}
\usepackage{makecell}
\usepackage[caption=false]{subfig} 
\usepackage[numbers,sort&compress]{natbib}
\usepackage{titletoc}
\usepackage{mathtools}

\newcommand{\reply}[1]{#1}

\begin{document}

\title{Adiabatic quantum state preparation in integrable models}

\author{Maximilian Lutz}
\affiliation{Max-Planck-Institut f{\"{u}}r Quantenoptik, Hans-Kopfermann-Str. 1, 85748 Garching, Germany}
\affiliation{Munich Center for Quantum Science and Technology (MCQST), Schellingstr. 4, D-80799 München, Germany}
\orcid{0000-0002-3805-3656}
\author{Lorenzo Piroli}
\affiliation{Dipartimento di Fisica e Astronomia, Università di Bologna and INFN, Sezione di Bologna, via Irnerio 46, I-40126 Bologna, Italy}
\orcid{0000-0002-0107-3338}
\author{Georgios Styliaris}
\affiliation{Max-Planck-Institut f{\"{u}}r Quantenoptik, Hans-Kopfermann-Str. 1, 85748 Garching, Germany}
\affiliation{Munich Center for Quantum Science and Technology (MCQST), Schellingstr. 4, D-80799 München, Germany}
\orcid{0000-0002-6809-8505}
\author{J.~Ignacio Cirac}
\affiliation{Max-Planck-Institut f{\"{u}}r Quantenoptik, Hans-Kopfermann-Str. 1, 85748 Garching, Germany}
\affiliation{Munich Center for Quantum Science and Technology (MCQST), Schellingstr. 4, D-80799 München, Germany}
\orcid{0000-0003-3359-1743}
\maketitle

\begin{abstract}
 We propose applying the adiabatic algorithm to prepare high-energy eigenstates of integrable models on a quantum computer. We first review the standard adiabatic algorithm to prepare ground states in each magnetization sector of the prototypical XXZ Heisenberg chain. Based on the thermodynamic Bethe ansatz, we show that the algorithm circuit depth is polynomial in the number of qubits $N$, outperforming previous methods explicitly relying on integrability. Next, we propose a protocol to prepare arbitrary eigenstates of integrable models that satisfy certain conditions. For a given target eigenstate, we construct a suitable parent Hamiltonian written in terms of a complete set of local conserved quantities. We propose using such Hamiltonians as an input for an adiabatic algorithm. After benchmarking this construction in the case of the non-interacting XY spin chain, where we can rigorously prove its efficiency, we apply it to prepare arbitrary eigenstates of the Richardson-Gaudin models. In this case, we provide numerical evidence that the circuit depth of our algorithm is polynomial in $N$ for all eigenstates, despite the models being interacting.
\end{abstract}


\section{Introduction}Integrable Hamiltonians provide a unique class of models in many-body physics, offering rich mathematical structure and rare analytical tractability~\cite{franchini2017introduction}. Even in these models, however, it is not known how to compute efficiently some important quantities. An example is given by correlation functions of highly excited states, where tensor network methods are unavailable~\cite{murg2012algebraic} and analytic computations are restricted to low-order correlators and special cases~\cite{pozsgay2011local, negro2013one, mestyan2014short, piroli2016multiparticle, pozsgay2017excited, bastianello2018exact}. In this context, quantum simulation~\cite{quantum_simulation_general_review} emerges as a potentially useful tool to go beyond the capabilities of classical methods. 

For non-interacting Hamiltonians, it is known that all eigenstates can be prepared on a quantum computer by quantum circuits scaling linearly in the number of qubits $N$~\cite{PhysRevLett.120.110501,PhysRevApplied.9.044036,PhysRevLett.133.230401}. This is very different from the eigenstates of generic Hamiltonians, for which the circuit depth is expected to scale exponentially in $N$. For \emph{interacting} integrable models, one would still expect that the structure of the wavefunctions is simpler than in the case of generic Hamiltonians, but there is currently no polynomial bound on the circuit depth required to prepare their eigenstates (with a few exceptions~\cite{ruiz2024efficienteigenstatepreparationintegrable}). For these reasons, the past few years have witnessed increasing efforts to develop quantum algorithms preparing eigenstates of integrable Hamiltonians, with the XXZ Heisenberg chain~\cite{korepin1997quantum} serving as a paradigmatic example. The existing algorithms are, however, variational~\cite{nepomechie2020bethe_publication,kannan2024quantum} or efficient only for a small subset of the eigenstates, generally requiring an amount of quantum or classical resources growing exponentially in $N$~\cite{ruiz2024bethe,sopena2022algebraic,van2021preparing,van2022preparing,li2022bethe,raveh2024deterministic_publication,raveh2024estimating_publication,mao2024toward,yeo2025reducingcircuitdepthquantum,sahu2024fractal}.

In this work, we propose to prepare eigenstates of integrable models by means of the adiabatic algorithm. The standard adiabatic algorithm is a powerful tool to prepare ground states of many-body systems, representing a common subroutine for quantum simulation~\cite{aspuru2005simulated,bauer2020quantum,PRXQuantum.5.020365}. It proceeds by slowly driving the system from the ground state of a trivial Hamiltonian to the target one, in such a way that the system wavefunction does not leave the instantaneous ground space. The efficiency of the protocol is dictated by the Hamiltonian gap. For this reason, the adiabatic algorithm is not usually considered to be useful to prepare highly excited states, as the corresponding energy eigenvalues are exponentially close to one another. Contrary to this expectation, we demonstrate that ideas based on the adiabatic algorithm can be useful to efficiently\reply{, in the sense of polynomial resource requirement scaling,} prepare eigenstates of (interacting) integrable models by a quantum computer.

\section{The adiabatic algorithm} We first review some fundamentals. In the adiabatic algorithm, the initial state of the system, $\ket{\psi(0)}$, is an eigenstate of a Hamiltonian $H(g(0))$ ~\cite{albash2018adiabatic}. Subsequently, the system is evolved by the time-dependent Hamiltonian $H(g(t))$, where $t \in [0, T]$ and $g(t)$ \reply{is a vector of parameters } \reply{which we take to be varied linearly like $g(t) = g^\textrm{initial} + (g^\textrm{final}-g^\textrm{initial}) t / T$.}
Assuming that the eigenstate is non-degenerate, that it is separated from the rest of the spectrum by a gap $\delta(g(t)) > 0$, \reply{that the Hamiltonian is twice continuously differentiable} and that
\reply{
\begin{align}\label{eq:adiabatic_theorem}
    T\!=\!\mathcal{O}\!\left(\!\max_{t}\!\frac{
     \Vert H^{(2)}(t)\Vert\!+\!\Vert H^{(1)}(t)\Vert}{\epsilon \delta^2(g(t))}
    \!+\! \frac{\Vert H^{(1)}(t)\Vert^2}{\epsilon \delta^3(g(t))}
    \!\right)\!,
\end{align}
}
the adiabatic theorem guarantees that the final state
\begin{equation}
    \ket{\psi(T)}= \mathcal{T} e^{-i\int_0^T H(g(t))dt}\ket{\psi(0)}\,,
\end{equation}
is an eigenstate of $H(g(T))$ to error $\epsilon$ ~\cite{jansen2007bounds,albash2018adiabatic}. Here, \reply{$\Vert\cdot\Vert$ denotes the operator norm, $\mathcal{T}\operatorname{exp}(\cdot)$ the time ordered exponential operator as well as $H^{(l)}(t) \coloneqq T^l \partial^l_{t} H(g(t))$ the derivatives of the Hamiltonian.\footnote{\reply{The power in $T$ stems from the fact that derivatives in the adiabatic theorem ($\partial_s$) need to be taken with regards to the dimensionless fraction of the total time $s \coloneqq t/T \in [0, 1]$.}}} 

The adiabatic algorithm can be implemented on a digital quantum computer by realizing the unitary time evolution \reply{under a time dependent Hamiltonian} as a quantum circuit (quantum simulation).
\reply{There is a variety of standard methods available for this task. Our overall interest focuses on constructing a quantum circuit with $D = \mathcal{O}(\mathrm{poly}(N))$ circuit depth scaling, as opposed to achieving an optimal power of the polynomial. We therefore consider product formulas for reasons of simplicity and relatively weak required conditions.
In particular, we assume a decomposition of the Hamiltonian into 1-sparse Pauli strings and a first order Trotter-Suzuki formula.}
\reply{Thus} let us consider a system of $N$ spins (qubits), with Hamiltonian $H(g) = \sum_{i=1}^S b_i(g) \vec{\sigma}_i$ where $\vec{\sigma}_i$ is an arbitrary tensor product of Pauli operators $\sigma^\alpha, \alpha=x,y,z$, \emph{i.e.} $H(g)$ has support on $S$ Pauli strings.
Then the circuit depth \reply{$D$} of \reply{its} quantum simulation is bounded by
\reply{\begin{equation}D = \mathcal{O}\left(N S \left(T \max_{\substack{l=0,1,2\\i, t}} \left(S \partial_t^l b_i(g(t))\right)^\frac{1}{1+l}\right)^{\frac{3}{2}} \varepsilon^{-\frac{1}{2}}\right),\end{equation}where $\varepsilon>0$ denotes an arbitrarily small error as measured by the operator norm difference between the ideal unitary time evolution and the product formula quantum circuit~\cite{wiebe2010higher,SM}. We do not discuss here additional required technical smoothness assumptions, as they will be satisfied in the following~\cite{SM}.
}

\reply{Summarizing the above}, we can identify \reply{sufficient} conditions under which the adiabatic algorithm is efficient, \emph{i.e.} the circuit depth scales as \reply{$D = \mathcal{O}(\mathrm{poly}(N))$}. This is the case if: $(i)$ the gap of the parent Hamiltonian $H(g)$ does not close faster than $\mathcal{O}(1/\mathrm{poly}(N))$ anywhere along the adiabatic path; $(ii)$ the number $S$ of Pauli strings is at most $\mathcal{O}(\mathrm{poly}(N))$ (\emph{i.e.} $H(g(t))$ is sparse); $(iii)$ the coefficients $b_i(g)$ and their derivatives are bounded by $\mathcal{O}(\mathrm{poly}(N))$.

\section{Integrable models and conservation laws} Integrable Hamiltonians are defined by the existence of an extensive number of conserved quantities (or charges), \emph{i.e.} operators $Q^{(k)}$ ($k = 1,\dots,N$ \reply{with $N$ system size})
such that 
\begin{align}
[H, Q^{(k)}] = 0\,, \!\!\quad\!\! [Q^{(k)}, Q^{(l)}] = 0 \!\quad\! \forall k,l=1,\dots,N,
\end{align}
where the $Q^{(k)}$ must satisfy certain sparsity constraints~\cite{Caux_2011}. A prominent example is given by so-called Yang-Baxter integrable models~\cite{korepin1997quantum}, which can be solved analytically by the Bethe ansatz formalism. The spectrum of integrable Hamiltonians is reminiscent of that of non-interacting fermionic systems: each eigenstate is associated with a set of quasiparticle excitations, whose momenta are called rapidities. Contrary to the noninteracting case, however, the rapidities are obtained as the solution of a nontrivial set of algebraic relations, known as the Bethe equations~\cite{korepin1997quantum}.

We will be interested in families of Hamiltonians $H(g)$, where $g$ are parameters that can be tuned preserving integrability. Further, we will assume that there is a value $g^\ast$ for which all eigenstates of $H(g^\ast)$ admit an efficient preparation protocol. Typically, this could occur at a point where the model becomes non-interacting.

\section{Adiabatic preparation of ground states} Before proceeding, we consider the application of the standard adiabatic algorithm for ground-state preparation. We exemplify it in the XXZ Heisenberg chain
\begin{align}
    H^{\mathrm{XXZ}}(\Delta)= \frac{-1}{4} \sum_{i=1}^N \sigma^x_i\sigma^x_{i+1} + \sigma^y_i\sigma^y_{i+1} + \Delta\sigma^z_i\sigma^z_{i+1},
    \label{eq:xxz_hamiltonian}
\end{align}
with $\sigma_{N+1}^\alpha = \sigma_1^\alpha$, which is a prototype for interacting integrable models.

The Hamiltonian~\eqref{eq:xxz_hamiltonian} commutes with total magnetization $M = \sum_{i=1}^N \sigma^z_i$ for all values of $\Delta$, so that $[M,\partial_\Delta H^{\rm XXZ}(\Delta)]=0$. Therefore, if we initialize the system in an eigenstate of $M$ for a given value of $\Delta$, it is easy to see that the eigenstates in different magnetization sectors are not coupled by the adiabatic evolution as we smoothly vary $\Delta$. In fact, it follows from the derivation of the adiabatic theorem~\cite{sakurai2020modern} that the gap in Eq.~\eqref{eq:adiabatic_theorem} can be computed only taking into account eigenstates with the same magnetization.

We now aim to prepare the lowest energy eigenstate of ~\eqref{eq:xxz_hamiltonian} in a sector with arbitrary magnetization $m$. We first prepare the ground state of this sector for the non-interacting spin chain $\Delta = 0$ using the algorithms put forward in Ref.~\cite{PhysRevLett.120.110501,PhysRevApplied.9.044036,PhysRevLett.133.230401}, whose circuit depth scaling is $\mathcal{O}(N)$.
Subsequently, we apply the adiabatic algorithm to reach an interacting point $\Delta\in ]0,1[$ \reply{in the gapless Luttinger-liquid like phase}.
$H^\mathrm{XXZ}$ can be simulated efficiently and the norms relevant to the adiabatic theorem (Eq.~\eqref{eq:adiabatic_theorem}) do not diverge.

The relevant gap for the adiabatic algorithm can be computed using the thermodynamic Bethe ansatz (TBA) formalism~\cite{takahashi2005thermodynamics}, which describes the spectrum of integrable systems in the large-$N$ limit.~\footnote{While the TBA is not mathematically rigorous, relying on the so-called string hypothesis, its predictions have been extensively tested against numerical computations and independent rigorous approaches when available, always finding perfect agreement~\cite{essler2005one}.}
The ground state in a sector of magnetization $m$ corresponds to the ground state of $H^\mathrm{XXZ}$ with an additional term $-h\sum_j \sigma^z_j$, where $h$ depends on $m$.
Within the TBA, each eigenstate of the low energy spectrum is labeled by three integers: the change in magnetization $N_M$, the number of ``backscattered quasiparticles'' $d$, and the number of particle-hole quasiparticle excitations $N^{\pm}$~\cite{franchini2017introduction}. 
The corresponding excitation energies take the form
$E\sim 2 \pi v_F \left(N_M^2/4 \mathcal{Z}^2 + \mathcal{Z}^2 d^2 + N^+ +N^-\right) N^{-1}$~\cite{franchini2017introduction}. The coefficients
$v_F$ and $\mathcal{Z}$ depend on $\Delta$ and $h$ and are solutions to a set of integral equations. Although they are not known analytically~\cite{franchini2017introduction}, they can be obtained numerically and are non-vanishing away from the quantum phase transition~\cite{SM}. As only eigenstates with the same magnetization as the ground state contribute to the gap, we take $N_M = 0$. This results in an overall gap scaling of $\mathcal{O}(1/N)$ for $0<\Delta<1$, as we have also verified by exact diagonalization numerics~\cite{SM}.

In conclusion, the adiabatic algorithm provides an efficient preparation protocol for the ground states in any arbitrary magnetization sector and exponentially improves on previous approaches that exploit integrability \cite{ruiz2024bethe,sopena2022algebraic,ruiz2024fbasis, van2021preparing,van2022preparing,li2022bethe, raveh2024deterministic_publication,raveh2024estimating_publication,mao2024toward,sahu2024fractal}.
However, unlike these prior methods, the above version of the adiabatic algorithm cannot be applied to prepare arbitrary high-energy eigenstates.

\section{Preparation of Arbitrary Excited States}
\label{sec:arbitary_eigenstate_prep_summary}
\reply{We now propose an algorithm that combines the adiabatic algorithm and integrability to go beyond ground states, and instead prepares arbitrary eigenstates in integrable models that satisfy certain conditions. 
We will first summarize the general concept of the construction as well as the complexity analysis, and then discuss their application to specific models in subsequent sections.}

Given \reply{an arbitrary} target eigenstate $\ket{v}$, we denote by $q_v^{(k)}$ the eigenvalue of the charge $Q^{k}$ associated with $\ket{v}$ and construct the parent Hamiltonian 
\begin{align}\label{eq:hv}
    H_v(g) = \sum_{k=1}^N (Q^{(k)}(g) - q^{(k)}_v(g))^2\,.
\end{align}
\reply{We will assume the completeness of the extensive set of integrals of motions $\{Q^{(k)}\}_{k=1}^N$~\footnote{The validity of this condition has been rigorously proven only in certain cases, but it is expected to hold generally~\cite{links2017completeness}.}; in situations where this only holds in the presence of extra symmetries, corresponding sectors thereof can be fixed by including additional commuting symmetry-resolving operators in an analogous manner.}
It is easy to see that $H_v$ is always positive-semidefinite and has $|v\rangle$ as its unique ground state. The square terms can be thought of as adding energy penalties for other eigenstates with different eigenvalues for at least one of the charges. If there is a value $g^\ast$ for which the ground state of $H_{v}(g^\ast)$ can be prepared efficiently, we can use $H_v(g)$ for an adiabatic algorithm to prepare $\ket{v}$ at the desired value of the Hamiltonian parameters.

\reply{Summarizing the complexity analysis stemming from the adiabatic theorem and quantum simulation,} the adiabatic evolution obtained by smoothly varying $g$ in Eq.~\eqref{eq:hv} implies the existence of an efficient quantum circuit under the following conditions: $(i)$ the operators $Q^{(k)}(g)$, and thus $H_v(g)$, can be expressed in terms of at most $S = \mathcal{O}(\mathrm{poly}(N))$ Pauli strings with weights of at most modulo $\mathcal{O}(\mathrm{poly}(N))$; $(ii)$ the gap of $H_v(g)$ closes slower than $\Omega(1/\operatorname{poly}(N))$ for all eigenstates $\ket{v}$. These circuits can additionally be constructed efficiently, if $(iii)$ $Q^{(k)}(g)$ and $q^{(k)}_v(g)$ can be computed classically in $\mathcal{O}(\mathrm{poly}(N))$ time for all $g$ in the adiabatic path, $|v\rangle$ and $k$.

\section{The XY chain} We first exemplify the proposed protocol in the non-interacting XY spin chain 
\begin{align}
   H^{\mathrm{XY}}\!=\!\frac{-1}{2}\!\sum_{j=1}^N\frac{1\!+\!\gamma}{2} \sigma_j^x \sigma_{j+1}^x\!+\!\frac{1\!-\!\gamma}{2} \sigma_j^y \sigma_{j+1}^y\!+\!h \sigma_j^z,
   \label{eq:hamiltonian_xy}
\end{align}
where $\sigma^\alpha_{N+1}\!=\!\sigma^\alpha_{1}$. We can take without loss of generality $\gamma \geq 0$, as the transformation $\gamma \rightarrow -\gamma$ can be implemented by a global $\pi/2$ rotation along the $z$-axis (which is a trivial quantum circuit). Similarly, we can assume $h\geq 0$. It is well-known that the model displays a quantum phase transition from an ordered to a disordered phase at $h=1$ for all $\gamma$~\cite{franchini2017introduction}, and that a Jordan-Wigner transformation maps $H^{XY}$ to a non-interacting fermionic Hamiltonian.
\reply{
With global parity operator $Z = \prod^N_{i=1} \sigma^z_{i}$, its diagonal form reads
$H^{\mathrm{XY}} = (1+Z)/2~H_{Z=+1} + (1-Z)/2~H_{Z=-1}$
with
\begin{equation}
    H_Z\!=\!\!\!\sum_{p \in \Gamma_Z}\!\!\!\sqrt{(\cos(p) - h)^2 + \gamma^2 \sin(p)^2} \left(\!c^\dagger_p c_p - \!\frac{1}{2}\!\right)\
    \label{eq:hamiltonian_xy_freefermions}
\end{equation}
after Fourier- and Bogoliubov transform.}
Here $c$ and $c^\dagger$ are fermionic ladder operators, while $\Gamma_Z$ is the set of allowed momenta in each parity sector. In particular $p = \frac{2\pi}{N} l$
with $l = \frac{1}{2}, \frac{3}{2}, \dots, N-\frac{1}{2}$ for $Z=+1$, while $l = 0, 1, \dots, N-1$ for $Z=-1$~\cite{franchini2017introduction}.

Based on the mapping to free fermions, all eigenstates of the XY model can be prepared by circuits of depth $\mathcal{O}(N)$ using the algorithms developed in Refs.~\cite{verstraete2009quantum,PhysRevApplied.9.044036,PhysRevLett.133.230401}.
We do not aim at improving these results. Rather, we invoke these algorithms for the initial state at the Ising point $\gamma = 1, h=0$ and subsequently demonstrate the preparation protocol by adiabatically sweeping to any other point in the phase as an illustrative example.
Let us first focus on $h<1$.

It can be seen immediately that the set of conserved operators $\{c_p^\dagger c_p | p = \frac{2\pi}{N}l, l=0,\frac{1}{2},\dots,N-1,N-\frac{1}{2}\}$ completely specify an eigenstate. 
Using this set, we construct the parent Hamiltonian
\reply{
\begin{align}\label{eq:parent_XY}
    H^{\mathrm{XY}}_v(\gamma, h) = (Z - z_v)^2 + \sum_{p\in \Gamma_{z_v}} (c^\dagger_p c_p - q_v^{(p)})^2.
\end{align}
Here $z_v$ is chosen as the desired parity of the state to be prepared and the inclusion of the corresponding first quadratic term serves the purpose of fixing the sector as previously described.
}
Now, it is immediate to see that the gap of this parent Hamiltonian is $\delta(\gamma, h)=1$, as two different eigenvectors differ in at least one mode occupation.

To proceed, we express $c^\dagger_p c_p$ in terms of the original spin degrees of freedom
\begin{multline}
    c^\dagger_p c_p
    = \sum_{j,l=1}^{N} \frac{e^{-ip(l-j)}}{N}
    \prod_{k=1}^{j-1}\sigma^z_k\\
    \begin{pmatrix}
        \sigma^x_j \\ \sigma^y_j
    \end{pmatrix}^T
    \begin{pmatrix}
      1 & i e^{2i\theta_p} \\
      -i e^{-2i\theta_p} & 1
    \end{pmatrix}
    \begin{pmatrix}
        \sigma^x_l \\ \sigma^y_l
    \end{pmatrix}
    \prod_{k=1}^{l-1}\sigma^z_k,
\label{eq:xy_number_operator_pauli_basis}
\end{multline}
where $\tan(2 \theta_p) = \frac{\gamma \sin(p)}{h-\cos(p)}$~\cite{franchini2017introduction}.  Eq.~\eqref{eq:xy_number_operator_pauli_basis} makes it explicit that these charges consist of $S = \mathcal{O}(N^2)$ Pauli strings and the modulo of their coefficients is $|b_i| = \mathcal{O}(1/N)$.
As the sum of squares of these terms, the parent Hamiltonian thus has \reply{at most} $S=\mathcal{O}(N^5)$ with coefficients $|b_i| = \mathcal{O}(1)$. 
In addition, the first and second derivatives of the parent-Hamiltonian matrix elements with respect to $\gamma$ and $h$ are system-size independent, only diverging at the phase boundaries of the XY model~\cite{SM}. \reply{If the adiabatic paths are chosen to linearly approach their final values inside the phase, this will result in only a constant factor in the final complexity dependent on how close the final $\gamma, h$ values are to the phase boundary~\cite{SM}.}
With the $c_p^\dagger c_p$ as above and since the $q^v_p$ are constant over the adiabatic path, the parent Hamiltonian can be computed efficiently.
In conclusion, we have verified that the Hamiltonian~\eqref{eq:parent_XY} satisfies all the hypotheses of the adiabatic theorem, implying that all eigenstates for $h<1$ can be prepared efficiently.
Similarly, when $h>1$, one can start from the classical Hamiltonian $h\rightarrow \infty$, and sweep $s \in [0, 1[$ after reparametrizing $s \leftrightarrow 1/h$.

\reply{We note that the parent Hamiltonian can be optimized to a potentially more practical form for quantum simulation by trading off the use of local integrals of motions in place of the $c^\dagger_p c_p$ against a smaller $\mathcal{O}(1/N)$ gap \cite{SM}. Similarly it is possible to rewrite the quadratic terms as a linear expression in the occupation number operators by exploiting the fermionic commutation relations \cite{SM}.}

\section{The Richardson-Gaudin models}We finally apply our construction for arbitrary eigenstate preparation of an interacting model. We consider the class of spin-$1/2$ XXX Richardson-Gaudin (RG) models~\cite{richardson1963,richardson1964, claeys2018richardson} $H^{\mathrm{RG}}=\sum_{k=1}^N \omega_k Q^{(k)}$, where
\begin{equation}
\label{eq:rg_Qk}
Q^{(k)}=\frac{\sigma_k^z}{2} + \frac{1}{2}+\frac{g}{4} \sum_{j \neq k}^N \frac{\vec{\sigma}_k \cdot \vec{\sigma}_j - 1}{\epsilon_k - \epsilon_j}\,.
\end{equation}
Here, $\omega_k$ and $\epsilon_k$ (with $\epsilon_k \neq \epsilon_j$ for $k \neq j$) are arbitrary parameters and $g$ is the interaction strength. The Hamiltonian commutes with the total magnetization $M$ and with the operators $Q^{(k)}$, which are mutually commuting $[Q^{(j)},Q^{(k)}]=0$~\cite{claeys2018richardson}. In fact, the operators $\{Q^{(k)}\}_{k=1}^N$ provide a complete set of conservation laws, \emph{i.e.} the set of their eigenvalues uniquely specify an Hamiltonian eigenstate~\cite{links2017completeness}. Setting $\omega_1 = 1$,  $\epsilon_1 = 0$ and $\omega_i = 0 $, $\epsilon_i = -e^{(i-2)/N}$ for $i\neq 1$, we obtain the central-spin model, which is relevant in the
study of quantum dots like the nitrogen-vacancy defect in diamond~\cite{nv_centers, claeys2018richardson}. 
In the RG models, each eigenstate $\ket{v}$ is labeled by the vector $\vec{q}_v = (q_v^{(1)}, \dots, q_v^{(N)})^T$ where $q_v^{(k)}$ is the eigenvalue of the charge $Q^{(k)}$ associated with $\ket{v}$. The set of all possible vectors $\vec{q}_v$ coincides with the set of solutions to the so-called quadratic Bethe equations
\begin{align}\label{eq:quadratic_bethe}
   {q^{(k)}}^2\!\!= q^{(k)}-\frac{g}{2}\sum_{j\neq k}^N \frac{q^{(k)}-q^{(j)}}{\epsilon_k - \epsilon_j},
\end{align}
with the additional constraint $\sum_{k} q^{(k)}\!=\!M/2+N/2$. Eqs.~\eqref{eq:quadratic_bethe} are equivalent to the traditional Bethe equations written in terms of the quasiparticle rapidities~\cite{claeys2018richardson}.
At $g=0$, the quadratic Bethe equations become trivial (${q^{(k)}}^2 = q^{(k)}$) and $\{\vec{q}_v\}_v$ coincides with the set of binary vectors.

For generic $g$, the solution to Eqs.~\eqref{eq:quadratic_bethe} can be found numerically by the following procedure: starting from $g=0$ and with a given binary vector, one iteratively increases $g$ by a small amount $\Delta g$ and solves Eqs.~\eqref{eq:quadratic_bethe} via a local optimization initialized with the solution of the previous step~\cite{PhysRevB.83.235124}.
As an improvement, the optimizer can also be initialized instead by computing a Taylor expansion at the previous solution, which reduces to solving linear systems of equations as for the quadratic Bethe equations~\cite{PhysRevB.83.235124}.

\begin{figure}
    \centering
    \includegraphics[width=\linewidth]{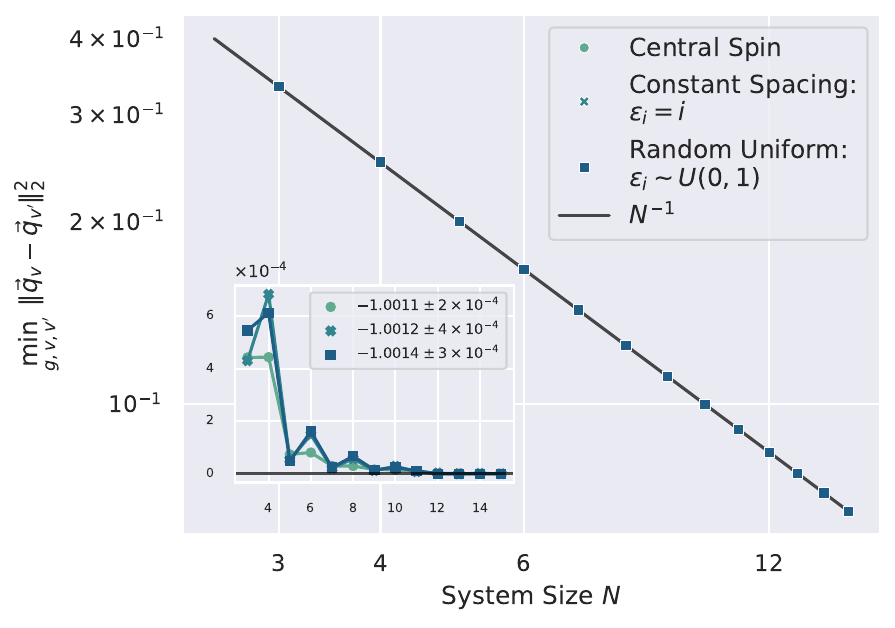}
    \caption{The RG quadratic Bethe equations are solved numerically for all eigenstates while dynamically adapting $\Delta g$, the pairwise $\Vert\cdot\Vert_2$-distance is computed and minimized over the adiabatic path $g\in[0,10]$ and all eigenstate pairs \cite{SM}.
This equals the minimal gap of $H^{RG}_v$ along the adiabatic path for all eigenstates $v$ and is shown against system size.
Inset shows deviations of the data point from the $N^{-1}$ curve shown in black; \reply{inset legend shows fitted exponents with standard deviation error}.
It can be seen that this fits the gap closing for all models.
}
\label{fig:rg_minimal_gap_scaling}
\end{figure}

Given a target eigenstate $\ket{v}$, we can follow our protocol and construct the parent Hamiltonian $H^{\mathrm{RG}}_v(g) = \Sigma_{k=1}^{N} (Q^{(k)}(g) - q_v^{(k)}(g))^2$, where the $q_v^{(k)}(g)$ are computed \reply{classically as described}. Note that the spectrum of $H^{\mathrm{RG}}_v$ can be expressed as the distance between the eigenvalue vectors $H^{\mathrm{RG}}_v |v^\prime\rangle = ||\vec{q}_v - \vec{q}_{v^\prime}||_2^2$.

For the adiabatic path, we vary $g$ after starting at $g=0$, where all the eigenstates are product states. By the definition of $Q^{(k)}$, $H^{\textrm{RG}}_v(g)$ has support on $S = \mathcal{O}(N^3)$ Pauli strings with coefficients scaling as $|b_i| = \mathcal{O}((g^2+1)/\min_{k,j}|\epsilon_k - \epsilon_j|^2)$, which for the central-spin model gives $|b_i| = \mathcal{O}(N)$.

Next, we study the gap of $H^{\textrm{RG}}_v(g)$. It is straightforward to show that states in different magnetization sectors ($M, M^\prime$) are separated by an $\mathcal{O}(1/N)$ energy
\begin{align}
\label{eq:rg_gap_different_magnetization}
\|\vec{q}_v - \vec{q}_{v^\prime}\|_2^2
\geq \frac{1}{N}\left|\sum_{k=1}^N q_v^{(k)} - q_{v^\prime}^{(k)}\right|^2
=\frac{(M-M^\prime)^2}{4N}.
\end{align}
\reply{
Moreover, in the limit $g\to 0$, the gap can be evaluated rigorously via standard perturbation theory: when
$g < \frac{\min_{k,j} |\epsilon_k - \epsilon_j|}{2 (2 + \sqrt{5}) N^2}$, we find that the gap is $\mathcal{O}(1)$~\cite{SM}.
Likewise, and with additional properties of the Gaudin magnet \cite{gaudin1976diagonalisation}, in the limit $g\rightarrow \infty$ we prove $\min_{v, v^\prime}
\|\vec{q}_v - \vec{q}_{v^\prime}\|_2^2
= 1/N$ \cite{SM}.
}
For states within the same magnetization sector \reply{and away from the perturbative regimes}, we were not able to rigorously bound the scaling of the gap, however, we performed a numerical analysis for different values of the parameters $\varepsilon_k$ \reply{with the technique described above}~\cite{SM}.
\reply{The minimization was carried out over all pairs of eigenvectors and for choices of $g$ connecting to both perturbative limits \cite{SM}}.
Our results are shown in Fig.~\ref{fig:rg_minimal_gap_scaling}, strongly supporting a $\mathcal{O}(1/N)$ scaling for increasing $N$. Notably, the numerics show a strictly monotonically decreasing behavior for $\min_{v, v^\prime}
\|\vec{q}_v - \vec{q}_{v^\prime}\|_2^2$ with $g$ between the weak and strong interaction limit values~\cite{SM}, thus further corroborating an overall $1/N$ lower bound for all $g$.

\reply{
    To our knowledge there is no rigorous result that guarantees or prohibits that Eqs.~\eqref{eq:quadratic_bethe} can be solved classically in polynomial time, and no runtime guarantee for the numerical technique employed here. This problem is equivalent to establishing the complexity of solving the Bethe equations as a more fundamental question about the Bethe ansatz and is beyond the scope of our work. A sufficient condition for the efficiency of the employed technique is that the $\Delta g$ required to remain in the local minima at any iteration does not vanish exponentially with increasing system size and the system of linear equations stays well-conditioned. While we are not able to provide conclusive evidence thereof as part of our numerics~\cite{SM}, the successful application of the algorithm here and in the numerical literature~\cite{PhysRevB.83.235124,johnson2020richardson,rg_numerical_method} provides evidence that efficient computation is possible in practice for at least many eigenstates. Thus we will take the efficient computability of $\vec{q}$ as a key assumption for the protocol.
}

In conjunction with the above, this implies the existence of quantum circuits for the preparation of all eigenstates with polynomial depth in $N$.

\section{The XXZ Model} We conclude by discussing why our protocol does not apply, as is, to the XXZ chain. The main bottleneck comes from the structure of the conserved charges. Indeed, it is known that the $k$-th charge of the XXZ model is written in terms of a number of Pauli strings $S=\mathcal{O}(N2^k)$~\cite{grabowski1995structure}, thus growing exponentially in $k$. In order to build the parent Hamiltonian $H^{XXZ}_v$ for all eigenstates, however, we need to include $\mathcal{O}(N)$ charges, so that $S$ would grow exponentially in \reply{system size} $N$, \reply{implying an $\mathcal{O}(\exp{(N)})$ circuit depth}.

A potential modification of our protocol would be to only include the first $\mathcal{O}(\log(N))$ charges in a parent Hamiltonian of the form $H^{\mathrm{XXZ,}\log(N)}_v = H^{\mathrm{XXZ}} + \sum_{k=1}^{\lfloor \log(N) \rfloor} (Q^{(k)}(g) - q_v^{(k)}(g))^2$. This way, however, only $\mathcal{O}(\mathrm{poly}(N))$ eigenstates could be targeted, which are the lowest energy states in sectors with $\log(N)$ fixed charges.
We note this has been considered from an entanglement scaling perspective \cite{alba2009entanglement}.
Making this construction precise, as well as trying to generalize the method to prepare all eigenstates of the XXZ model, are interesting directions for future research.

\section{Outlook} Our work raises several questions. First, it would be important to rigorously prove the scaling for the parent Hamiltonian that we found numerically in the RG models. A possible way to do so would be to include bounds on root separation of algebraic equations, perhaps by means of Smales $\alpha$-theory~\cite{SM,blum1998complexity}. \reply{Next, our work further motivates future work on classical computational complexity aspects of the Bethe ansatz from the quantum algorithms perspective. Moreover,} it would be very interesting to characterize the models for which our protocol can be applied  (possibly, with suitable modifications), such as potentially models of ``free fermions in disguise''~\cite{fendley2019free}.

\section{Acknowledgments}
The authors thank Mingru Yang, Sachin Teli, Yuhan Liu, Zhiyuan Wang and Reinis Irmejs for helpful discussions.
M.L., G.S. and J.I.C. acknowledge support from the German Federal Ministry of Education and Research (BMBF) through the funded project ALMANAQC, grant number 13N17236 within the research program ``Quantum Systems''. The research is partly funded by THEQUCO as part of the Munich Quantum Valley, which is supported by the Bavarian state government with funds from the Hightech Agenda Bayern Plus.
The work of L.P. is funded by the European Union (ERC, QUANTHEM, 101114881).
Views and opinions expressed are however those of the author(s) only and do not necessarily reflect those of the European Union or the European Research Council Executive Agency.
Neither the European Union nor the granting authority can be held responsible for them.
Exact diagonalization calculations were performed using QuSpin~\cite{weinberg2017quspin}.\\

\textbf{Data Availability Statement.---}
The source code and the data have been deposited in the Zenodo public folder \cite{zenodo}. 

\bibliographystyle{plainnat}
\bibliography{refs}

\onecolumngrid
\newpage

\appendix
\setcounter{equation}{0}
\setcounter{figure}{0}
\renewcommand{\thetable}{S\arabic{table}}
\renewcommand{\theequation}{S\thesection.\arabic{equation}}
\renewcommand{\thefigure}{S\arabic{figure}}
\setcounter{secnumdepth}{2}

\begin{center}
    {\Large \bf Supplemental Material}
\end{center}

Here we provide additional details about the results stated in the main text.

\section*{Contents}  
\startcontents[appendix]  
\printcontents[appendix]{}{0}{\setcounter{tocdepth}{2}}

\section{XY Model Details}
\label{sec:xy_local_ioms}
Here we provide more context for the parent Hamiltonian construction for the XY model.
We first briefly review the mapping of the XY spin chain to a free fermion Hamiltonian~\cite{franchini2017introduction} to fix the conventions, which proceeds first by transforming the XY Hamiltonian (Eq.~\ref{eq:hamiltonian_xy}) to 
\begin{align}
H^\textrm{XY}=-\frac{1}{2} \sum_{j=1}^{N-1}\left(a_j^{\dagger} a_{j+1}+a_{j+1}^{\dagger} a_j+\gamma a_j^{\dagger} a_{j+1}^{\dagger}+\gamma a_{j+1} a_j\right)+h \sum_{j=1}^N a_j^{\dagger} a_j-\frac{h N}{2} +\frac{Z}{2}\left(a_N^{\dagger} a_1+a_1^{\dagger} a_N+\gamma a_N^{\dagger} a_1^{\dagger}+\gamma a_1 a_N\right),
\end{align}
via Jordan-Wigner transformation
$(\sigma_j^+ = \prod_{l=1}^{j-1}(1-2 a_l^\dagger a_l)a_l,\sigma_j^- = \prod_{l=1}^{j-1}(1-2 a_l^\dagger a_l)a^\dagger_l,
\sigma_j^z = 1-2 a_j^\dagger a_j)$
with fermionic creation/annihilation operators $a^\dagger/a$~\cite{franchini2017introduction}.
The periodic ($Z = -1$) / antiperiodic ($Z=+1$) boundary conditions can be implemented by going to parity sectors: $H^\textrm{XY} = \frac{1+Z}{2} H_{Z=+1} + \frac{1-Z}{2} H_{Z=-1}$, where
\begin{align}
    H_{Z=\pm1}=-\frac{1}{2} \sum_{j=1}^N\left(a_j^{( \pm) \dagger} a_{j+1}^{( \pm)}+a_{j+1}^{( \pm) \dagger} a_j^{( \pm)}+\gamma a_j^{( \pm) \dagger} a_{j+1}^{( \pm) \dagger}+\gamma a_{j+1}^{( \pm)} a_j^{( \pm)}-2 h a_j^{( \pm) \dagger} a_j^{( \pm)}\right)-\frac{h N}{2}
\end{align}
and the particle number is restricted to be even / odd for $Z=\pm1$~\cite{franchini2017introduction}.
We then Fourier transform $a_p = N^{-1/2} \sum_{j=1}^N \exp{(-i p j)} a_j^{(\pm)}$ ($\{a_p, a_q^\dagger\}=\delta_{p,q}$) with momenta $\Gamma_Z$ equal to $p = \frac{2 \pi l}{N}$ with $l = \frac{1}{2}, \frac{3}{2}, \dots, N-\frac{1}{2}$/$l = 0, 1, \dots, N-1$ for $Z=+1/-1$ (to respect the boundary conditions) to arrive at~\cite{franchini2017introduction}
\begin{align}
    H_{Z} = \frac{1}{2} \sum_{p\in \Gamma_Z}
    \begin{pmatrix}
        a_p^\dagger & a_{-p}
    \end{pmatrix}
    \begin{pmatrix}
        e^{i \pi / 4} & 0\\
        0 & e^{-i \pi / 4} \\
    \end{pmatrix}
    \begin{pmatrix}
        h-\cos(p) & -\gamma \sin(p)\\
        -\gamma \sin(p) & \cos(p)-h \\
    \end{pmatrix}
    \begin{pmatrix}
        e^{-i \pi / 4} & 0\\
        0 & e^{i \pi / 4} \\
    \end{pmatrix}
    \begin{pmatrix}
        a_p^\dagger \\ a_{-p}^\dagger
    \end{pmatrix}.
\end{align}
Finally a Bogoliubov transform $
\begin{pmatrix}
    a_p \\ a_{-p}^\dagger
\end{pmatrix}
=
\begin{pmatrix}
    e^{i \pi / 4} & 0\\
    0 & e^{-i \pi / 4} \\
\end{pmatrix}
\begin{pmatrix}
    \cos(\theta_p) & \sin(\theta_p) \\
     -\sin(\theta_p) & \cos(\theta_p) \\
\end{pmatrix}
\begin{pmatrix}
    c_p \\ c_{-p}^\dagger
\end{pmatrix},
$
where the Bogoliubov angle is defined by $\tan(2\theta_p) = \frac{\gamma \sin(p)}{h-\cos(p)}$, brings the Hamiltonian into free fermion form $H_Z = \sum_{p\in \Gamma_Z} \epsilon(p) (c_p^\dagger c_p - 1/2)$ with dispersion $\epsilon(p) = \sqrt{(h-\cos(p))^2 + \gamma^2\sin(p)^2}$~\cite{franchini2017introduction}.

For the norms of the derivatives along the adiabatic path, we can w.l.o.g. consider first linearly interpolating to the desired $h$ and subsequently linearly to the desired $\gamma$, as one can think of decomposing the adiabatic algorithm into two adiabatic sweeps.
The derivatives of the parent Hamiltonian then take the form $\partial_g H^\mathrm{XY}_\mathrm{v} = \sum_{p, Z} 2(c^\dagger_p c_p - q^{(p)}_v)\partial_g c^\dagger_p c_p$ and $\partial^2_g H^\mathrm{XY}_\mathrm{v} = \sum_{p, Z} 2(c^\dagger_p c_p - q^{(p)}_v)\partial^2_g c^\dagger_p c_p + 2 (\partial^2_g c^\dagger_p c_p)^2$, which implies we only need to consider the terms $\partial_g c^\dagger_p c_p, \partial^2_g c^\dagger_p c_p$ up to a $\mathcal{O}(\mathrm{poly}(N))$ factor.
For these one can consider Eq.~\ref{eq:xy_number_operator_pauli_basis}, where similarly up to a polynomial factor we only need consider the derivative along $\gamma$/$h$ applied to the matrix element containing the Bogoliubov angle and thus
\begin{align}
& \partial_\gamma e^{2i\theta_p} = \frac{i \sin (p) (h-\cos (p))}{(h-i \gamma  \sin (p)-\cos (p)) \epsilon(p)}
\quad
\partial^2_\gamma e^{2i\theta_p} = \frac{\sin ^2(p) (\cos (p)-h) (h+2 i \gamma  \sin (p)-\cos (p))}{(h+i \gamma \sin (p)-\cos (p)) (-h+i \gamma  \sin (p)+\cos (p))^2 \epsilon(p)}\\
&\partial_h e^{2i\theta_p} = \frac{\gamma  \sin (p)}{(\gamma  \sin (p)+i (h-\cos (p))) \epsilon(p)}
\quad
\partial^2_h e^{2i\theta_p} = -\frac{i \gamma  \sin (p) (-2 h-i \gamma  \sin (p)+2 \cos (p))}{(h+i \gamma \sin (p)-\cos (p)) (-h+i \gamma  \sin (p)+\cos (p))^2 \epsilon(p)} 
\end{align}
and
\begin{align}
& \partial_\gamma e^{-2i\theta_p} = \frac{i \sin (p) (\cos (p)-h)}{(h+i \gamma  \sin (p)-\cos (p)) \epsilon(p)}
\quad
\partial^2_\gamma e^{-2i\theta_p} = \frac{\sin ^2(p) (\cos (p)-h) (h-2 i \gamma  \sin (p)-\cos (p))}{(h-i \gamma 
   \sin (p)-\cos (p)) (h+i \gamma  \sin (p)-\cos (p))^2 \epsilon(p)}
\\
&\partial_h e^{-2i\theta_p} = \frac{\gamma  \sin (p)}{(-i h+\gamma  \sin (p)+i \cos (p)) \epsilon(p)}
\quad
\partial^2_h e^{-2i\theta_p} = \frac{i \gamma  \sin (p) (-2 h+i \gamma  \sin (p)+2 \cos (p))}{(h-i \gamma \sin (p)-\cos (p)) (h+i \gamma  \sin (p)-\cos (p))^2 \epsilon(p)},
\end{align}
which show that the coefficients only diverge at the phase boundaries, i.e. $|h|=1$ or $|h|<1, \gamma=0$.
This is sufficient for both the assumptions of the adiabatic theorem \cite{albash2018adiabatic} as well as the smoothness condition of the first-order product formula for a time dependent Hamiltonian \cite{wiebe2010higher}.

Here we comment on the number of IOMs $c_p^\dagger c_p, p\in\Gamma_Z$ and the dimension of the Hilbert space, which is $2^N$.
Both for $Z=+1$ and $Z=-1$ there are $N$ IOMs, however the naive estimate that this gives a state space of dimension $2^{2N}$ neglects the particle number constraints in the parity sectors.
More precisely, the $Z=+1/-1$ sector require even / odd particle number, the dimension of the spaces is therefore equal to the number of ways to choose an even / odd number of momenta from $\Gamma_Z$, which corresponds to $\binom{N}{0}+\binom{N}{2}+\binom{N}{4}+\dots = \binom{N}{1}+\binom{N}{3}+\binom{N}{5}+\dots = 2^{N-1}$ and thus exactly to the Hilbert space dimensions of each of the parity sectors.

\reply{Moreover we note, that the parent Hamiltonian can be simplified by the shift by a constant of $(c^\dagger_p c_p - q_v^{(p)})^2 = \mathrm{const.} + c^\dagger_p c_p (1-2q_v^{(p)})$.}

\section{XY Model Local Integrals of Motion}
In the algebraic Bethe ansatz, IOMs are constructed as log-derivatives of the transfer matrix and as such have a locality structure - specifically $Q^{(k)}$ are translation invariant sums of at most $k$-local operators.
In the main text it was mentioned, that this behavior can be mimicked in the non-interacting XY case with an alternative set of IOMs.

\cite[App. C]{fagotti2013reduced} constructs such quantities for the bulk Ising model, which directly generalizes to the case of the XY model, when (using the notation thereof)
$
    \mathcal{Y}_0 = \frac{i}{2} \begin{pmatrix} & h \\ -h &\end{pmatrix},
    \mathcal{Y}_1 = \frac{i}{2} \begin{pmatrix} & -\frac{1-\gamma}{2} \\ \frac{1+\gamma}{2} &\end{pmatrix}
$
and thus $
    Y_k = \frac{-i J}{2} \begin{pmatrix} 0 & h - \frac{1-\gamma}{2} e^{-2\pi i k/L} - \frac{1+\gamma}{2} e^{2 \pi i k / L} 
    \\ -h + \frac{1+\gamma}{2} e^{-2\pi i k/L} + \frac{1-\gamma}{2} e^{2\pi i k/L} & 0\end{pmatrix}
$.
As a $2\times2$ traceless matrix, solution~\cite[(C7)]{fagotti2013reduced} holds analogously with Hamiltonian range $r_H = 1$.
Note however, that the construction of~\cite[App. C]{fagotti2013reduced} is for the infinite system limit and omits the boundary conditions, and indeed does not strictly work for the antiperiodic case (\cite[Eq. (C2)]{fagotti2013reduced} gains an additional $\pm$ in the uppermost-rightmost/lowest-leftmost matrix block element, breaking the necessary block circulant structure).
It, however, immediately suggests the following ansatz for the LIOMs:
\begin{align}
    Q_\textrm{loc}^{(k,Z)} =\!\sum_{p \in \Gamma_Z} \cos(p k) \sqrt{(\cos(p) - h)^2 + \gamma^2 \sin(p)^2} c_p^\dagger c_p.
\end{align}
Using Eq.~\ref{eq:xy_number_operator_pauli_basis} this gives
\begin{align}
    Q_\textrm{loc}^{(k,Z)}
    = \sum_{j,l=1}^{N} \frac{1}{N}
    \prod_{k=1}^{j-1}\sigma^z_k
    \begin{pmatrix}
        \sigma^x_j \\ \sigma^y_j
    \end{pmatrix}^T
    \sum_{p \in \Gamma_Z} \cos(p k) 
    \epsilon(p)
    e^{-ip(l-j)}
    \begin{pmatrix}
      1 & i e^{2i\theta_p} \\
      -i e^{-2i\theta_p} & 1
    \end{pmatrix}
    \begin{pmatrix}
        \sigma^x_l \\ \sigma^y_l
    \end{pmatrix}
    \prod_{k=1}^{l-1}\sigma^z_k.
\end{align}
For the $1$ matrix elements, for all $j\neq l$, the terms cancel as $\cos(p k) \epsilon(p)$ is invariant under $p \leftrightarrow 2\pi-p$, while the sign in $e^{-ip(l-j)}$ changes and the Pauli strings anticommute and we sum over all $j,k$ and $p$.
If $j=l$ the Pauli strings are identical and produce the identity with a factor.
For the upper right hand matrix element, we consider
\begin{align}
\sum_{p \in \Gamma_Z} \cos(p k) \epsilon(p) e^{-ip(l-j)} i e^{2i\theta_p}
=& \sum_{p \in \Gamma_Z} \cos(p k)  e^{-ip(l-j)} i (h - \cos(p) + i\gamma \sin(p))\\
=& N\left(\frac{-1}{4}(\delta_{j-l-k-1} + \delta_{j-l-k+1}+\delta_{j-l+k-1}+\delta_{j-l+k+1})
+\frac{h}{2}(\delta_{j-l-k} + \delta_{j-l+k})\right.
\\&\left.+\frac{\gamma}{2}(-\delta_{j-l-k-1} +\delta_{j-l-k+1}-\delta_{j-l+k-1}+\delta_{j-l+k+1})\right),
\end{align}
where in the first line equality we use the defining relation for the Bogoliubov angle and in the second line we define for $\alpha \in \mathbb{Z}$
\begin{align}
\sum_{p\in \Gamma_Z} e^{ip\alpha} =
\delta_\alpha := \left\{\begin{array}{lr}e^{i\pi \alpha /N}\delta_{\alpha,0}&\text{for } Z=+1\\ \delta_{\alpha,0}&\text{for } Z=-1 \end{array}.\right.
\end{align}
The other two matrix elements are analogous.
From this it can be seen, that the ansatz above for $Q_\textrm{loc}^{(k,Z)}$ contains only Pauli strings that are at most $k+1$-local, as the $\delta$-functions fix $|j-l| = k \pm 1$ or $|j-l| = k$.

As mentioned in the main text, a parent Hamiltonian can be constructed from these local IOMs as well in an analogous manner
\begin{equation}
    H^{\textrm{XY}}_\textrm{P,loc} = \sum_{z = \pm 1}\sum_{k=1}^N \left(\sum_{p\in \Gamma_z} \cos(p k) \epsilon(p) c^\dagger_p c_p - q_v^{(p,\textrm{loc})}\right)^2.
\end{equation}
This Hamiltonian is not gapped by $1$ anymore, as was evidently the case for $H^\textrm{XY}_\textrm{P}$.
A way to formulate a sufficient condition for the parent Hamiltonian gap would be to define two matrices $A^\pm$, which map $N$-dimensional binary vectors, whose entries represent the $c^\dagger_p c_p$ in the even / odd sector, to vectors containing the resulting values of the local integrals of motion $Q_\textrm{loc}^{(k,Z)}$.
These take the form
\begin{align}
    A^+_{n,l} := \cos\left(\frac{(n+1) 2\pi(l+1/2)}{N}\right)\epsilon\left(\frac{2(l+1/2)\pi}{N}\right)
    \quad
    \quad
    A^-_{n,l} := \cos\left(\frac{(n+1) 2\pi l}{N}\right)\epsilon\left(\frac{2\pi l}{N}\right)
\end{align}
with $n,l=0,...,N-1$, i.e. $A^\pm$ are quadratic.
Then if the $c^\dagger_p c_p$ eigenvalues of the eigenstate that should be prepared corresponds to the binary vectors $b_1^\pm$, let us consider the value of the parent Hamiltonian on any other eigenstate $|o\rangle$ corresponding to $b_2^\pm$, which is not identical $b_1^+ \neq b_2^+ \lor b_1^- \neq b_2^-$:
\begin{align}
    H^{\textrm{XY}}_\textrm{P,loc} |o\rangle = ||A^+ b^+_1 - A^+ b^+_2||^2_2 + ||A^- b^-_1 - A^- b^-_2||^2_2 = \sum_{z=\pm} ||A^z (b^z_1 - b^z_2)||^2_2,
\end{align}
where the two norm stems from the sum of squares form of the Hamiltonian.
This can then be lower bounded by the matrix norm inequality
$
    \sum_{z=\pm} ||A^z (b^z_1 - b^z_2)||^2_2 \geq \sum_{z=\pm}||b^z_1 - b^z_2||_2^2 \min \sigma^2(A^z),
$
where $\min \sigma()$ denotes the minimal square singular value.
As $b^\pm_1$ and $b^\pm_2$ denote different eigenstates, they differ in at least one element, thus $||b^\pm_1-b^\pm_2||_2^2\geq 1 = O(1)$ for at least one of $+$ and $-$.
For the entire parent Hamiltonian to be gapped by at least an inverse polynomial value, it is thus sufficient to prove that $\min \sigma^2(A^\pm) \geq O(1/\text{poly}(N))$.
By the orthogonality of cosine functions, $({A^+}^\dagger A^+)_{n,l} = \frac{N}{2} \epsilon\left(\frac{2\pi(n+1/2)}{N})\right)^2 \delta_{n,l}$ and $({A^-}^\dagger A^-)_{n,l} = \frac{N}{2} \epsilon\left(\frac{2\pi n}{N}\right)^2 \delta_{n,l}$.
As these are diagonal matrices, their eigenvalues and thus the squared singular values become trivial.
The minimal squared singular value then corresponds to $\frac{N}{2} \min_{j\in 0,\frac{1}{2},1,\dots,N-1,N-\frac{1}{2}} \epsilon(\frac{2 \pi j}{N})^2$.
This corresponds to the minimum when probing $\epsilon(p)^2$ at different equally spaced values, which can be lower bounded by the continuum minimum of the $\epsilon^2(p) = (\cos(p) - h)^2 + \gamma^2 \sin(p)^2$ function:
\begin{equation}
\min_{p\in[0, 2\pi[} \epsilon^2(p) = 
 \left\{ 
\begin{array}{cc}
     (h-1)^2 & \left(|\gamma| <1\land \gamma ^2+h>1\right)\lor (h>0\land |\gamma| \geq 1) \\
     (h+1)^2 & (h<0\land |\gamma| = 1)\lor \left(|\gamma| <1\land h+1\leq \gamma ^2\right)\lor (|\gamma| > 1 \land h\leq 0) \\
     \gamma ^2 & h=0\land |\gamma| \leq 1 \\
     \frac{\gamma ^2 \left(\gamma ^2+h^2-1\right)}{\gamma ^2-1} & \text{else} \\
\end{array} \right\}
\end{equation}
which is $0$ exactly only at the phase boundaries.
In summary, this proves that $$H^{\text{LIOM}}_P |o\rangle \geq O(1) N \left(\min_{p\in[-\pi, \pi[}{\epsilon^2(p)}\right) \in O(N).$$
The orthogonality causes $A^\pm$ to be invertible if the singular values do not become zero, i.e. if we are not at a phase boundary of the XY model.
As the mode occupations $c^\dagger_p c_p$ are clearly a complete set of eigenvalues, the same is true for the set of LIOMs defined this way.

\section{XXZ ED Details}
\begin{figure}
    \centering
    \includegraphics[width=.9\linewidth]{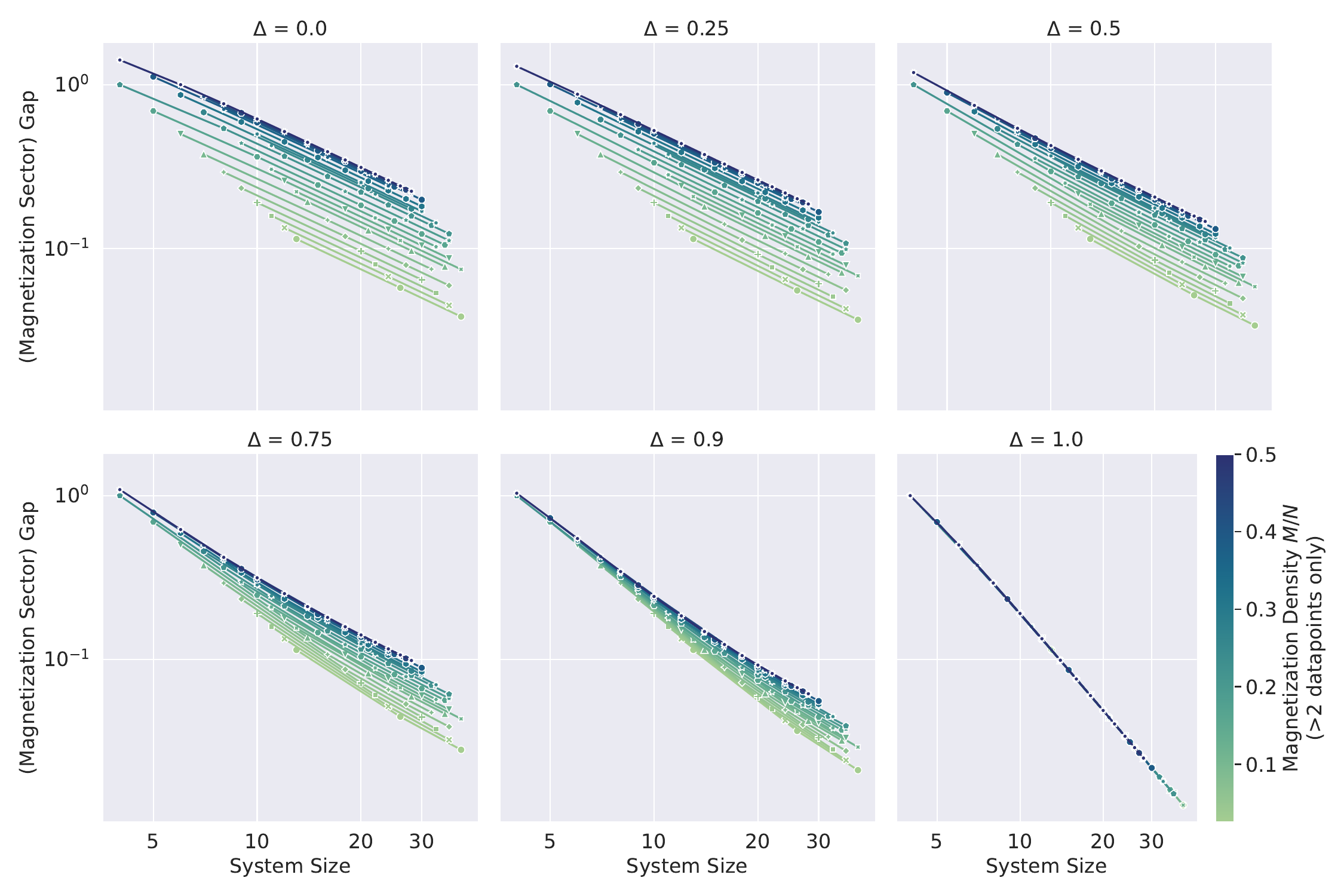}
    \caption{Exact diagonalization results for the gap above magnetization sector lowest-energy states in the XXZ model.
    }
    \label{fig:xxz_gap_scaling}
\end{figure}
\begin{figure}
\label{sec:si_xxz_ed}
\centering
\includegraphics[width=\linewidth]{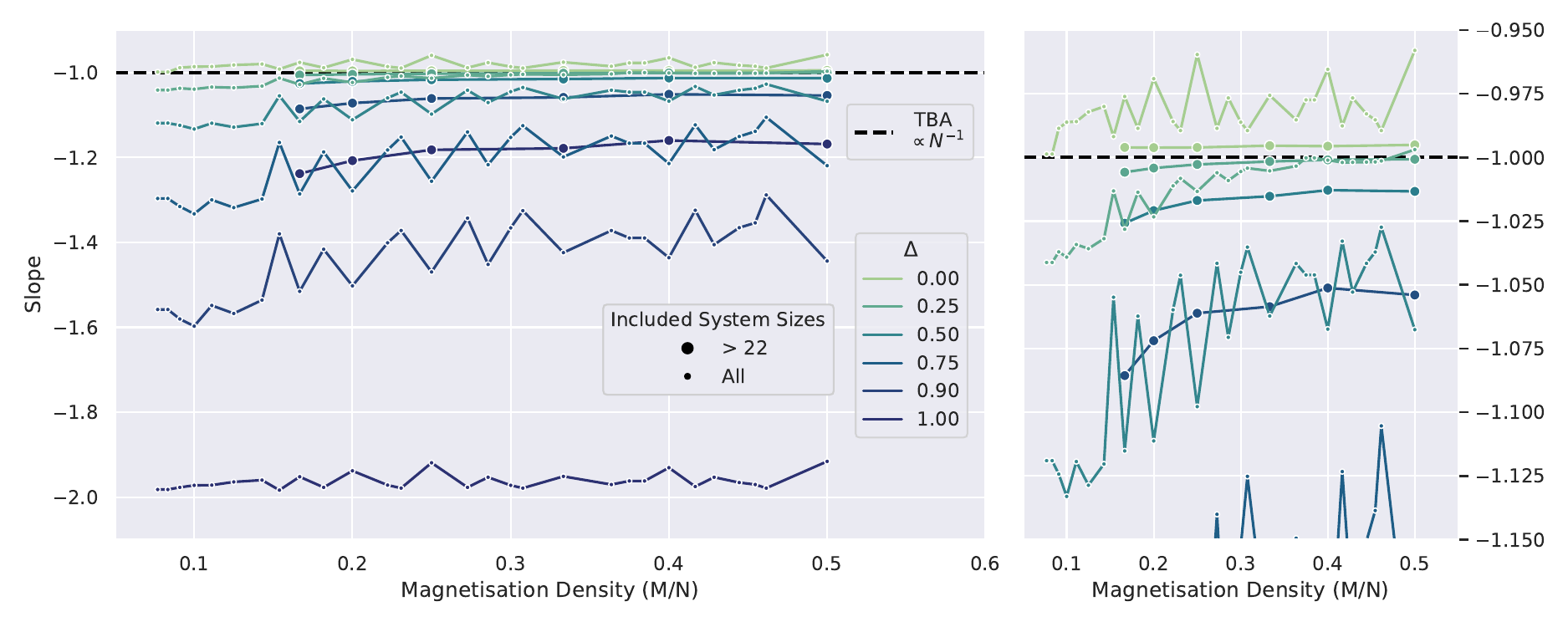}
\caption{
Extrapolated slopes for data shown in Fig.~\ref{fig:xxz_gap_scaling}.
Large dots corresponds to system sizes restricted to $N > 22$, smaller dots include all smaller system sizes.
Black line corresponds to thermodynamic Bethe ansatz $N^{-1}$ gap scaling.
The right hand figure is a magnification for y-values around $-1$.
}
\label{fig:xxz_slopes_supplementary}
\end{figure}
The gap of the Hamiltonian in the magnetization sectors is of crucial importance to the preparation efficiency for magnetization sector lowest-energy states in the XXZ model discussed in the main text.
To verify the large system size limit thermodynamic Bethe ansatz results, exact diagonalization computations were carried out using QuSpin \cite{weinberg2017quspin}.
The results are shown in Fig.~\ref{fig:xxz_gap_scaling}.
The slopes of the fits in the log-log plot were extracted as the power of the gap-system size relationship and are shown in Fig.~\ref{fig:xxz_slopes_supplementary}.
Only magnetization densities with $\geq 3$ data points were used, to avoid unreasonable fitting errors.
The ED small-system size results agree well with the large system limit TBA scaling.
The remaining deviations can be interpreted to stem from finite-size effects - which can be seen by comparing all system sizes and $N>22$ in Fig.~\ref{fig:xxz_slopes_supplementary}.
It appears that at $\Delta = 1$, i.e. the phase transition, the gap scales according to $N^{-2}$.
In a finite size system, the scaling of the gap is going to deviate from the TBA scaling of $N^{-1}$ even in the paramagnetic phase.
Indeed this is the case by approaching the $N^{-2}$ scaling, more so for $\Delta$ values closer to $1$ (i.e. closer to the phase transition point), as well as for smaller system sizes more than for larger ones (i.e. stronger finite size effects).

\section{XXZ Thermodynamic Bethe Ansatz Details}
\label{sec:si_xxz_tba}
\begin{figure}
\subfloat[
$v_F$: analytical limit is $\frac{\pi \sin(\arccos(-\Delta))}{2 \arccos(-\Delta)}$~\cite{franchini2017introduction}.
Vanishes at $\Delta = 1$; numerics is unstable very close to this point, which explains spurious negative values on colorbar.
]{\includegraphics[width=.32\linewidth]{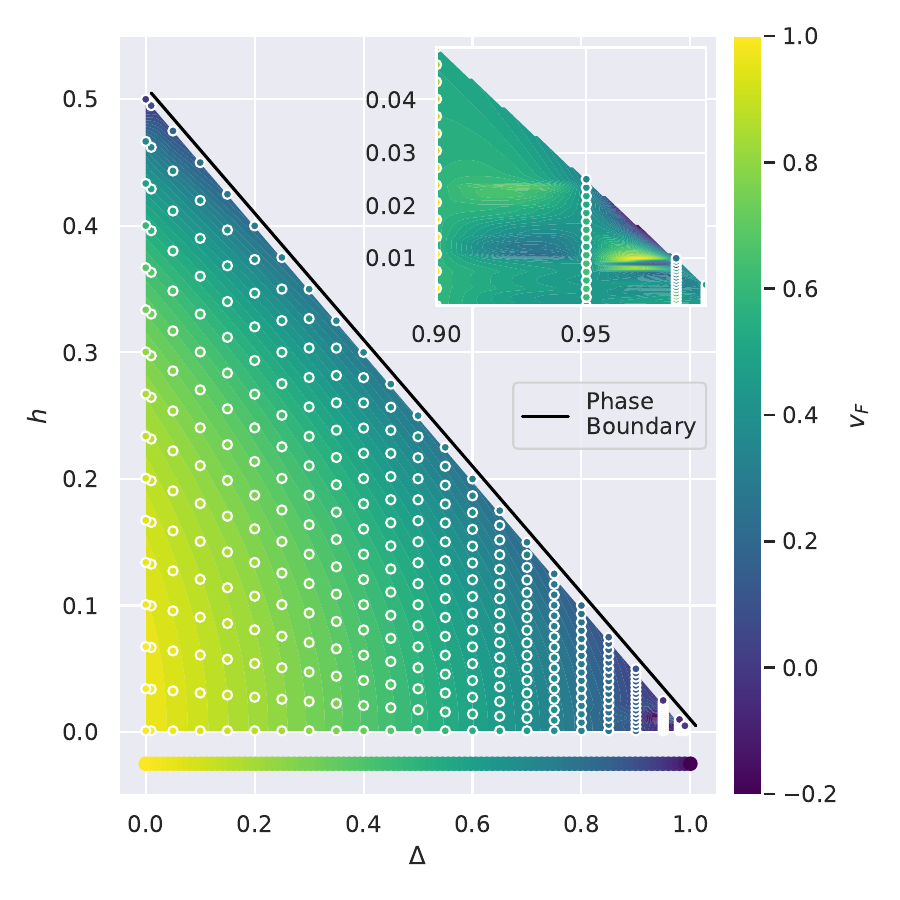}}
\hspace{2em}
\subfloat[
$\mathcal{Z}^2$: analytical limit is $\frac{\pi}{2 \left( \pi - \arccos(-\Delta) \right)}$~\cite{franchini2017introduction}.
Diverges at $\Delta = 1$, does not vanish anywhere in the phase.
]{\includegraphics[width=.32\linewidth]{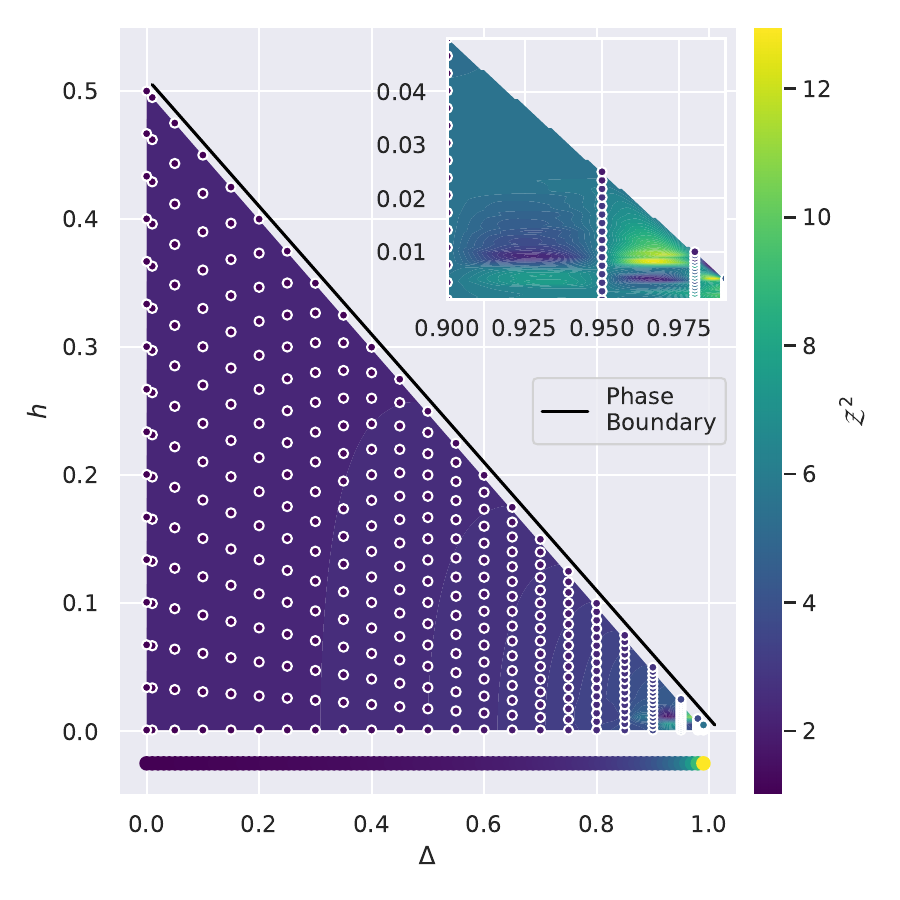}}
\subfloat[
$M/N$
]{\includegraphics[width=.32\linewidth]{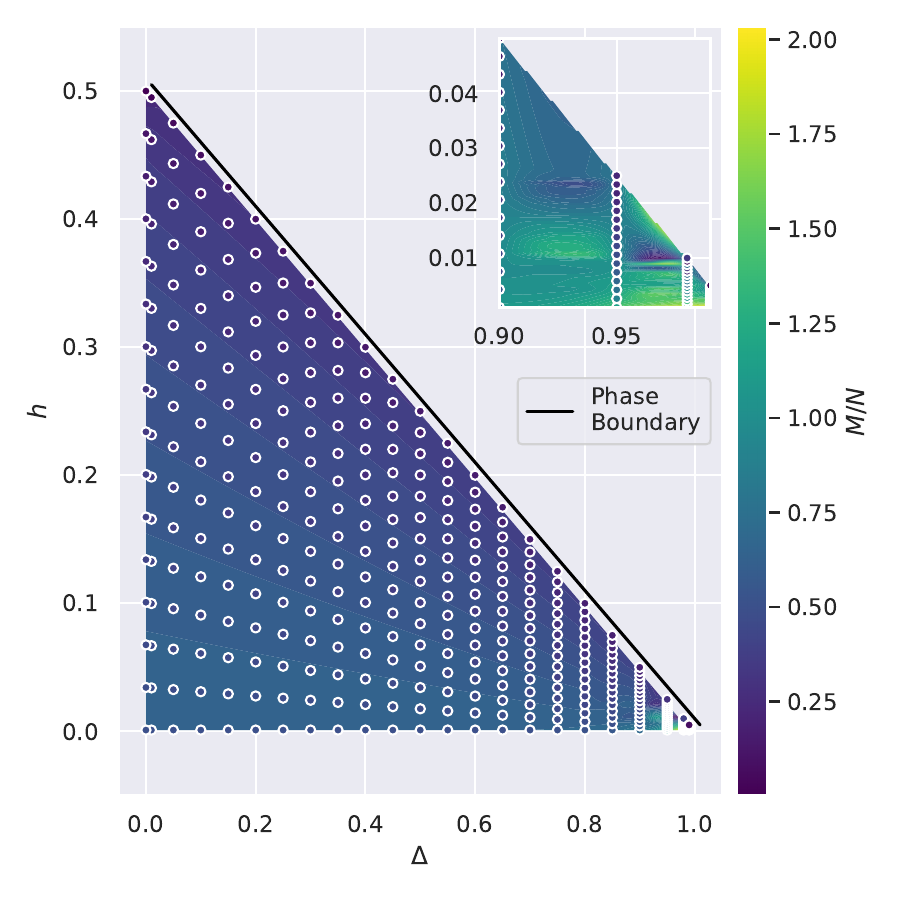}}
\caption[]{
Numerical results of iterated solution of integral equations for different $(h, \Delta)$ as circled dots; background color shows cubic interpolation for entire phase.
Black line marks phase boundary $h_\textrm{crit} = \frac{1-\Delta}{2}$ (slight offset) from para- to ferromagnetic phase.
The phase is symmetric for $h \rightarrow -h$, so only the positive part is shown.
The inset shows a magnification of the plot near $\Delta = 1$, only at which point the numerical solution can seen to be unstable.
Bottom lines show analytic solution for limit case $h=0$, it agrees with the numerical data.
}
\label{fig:tba_background}
\end{figure}
\begin{figure}
    \centering
    \includegraphics[width=.7\linewidth]{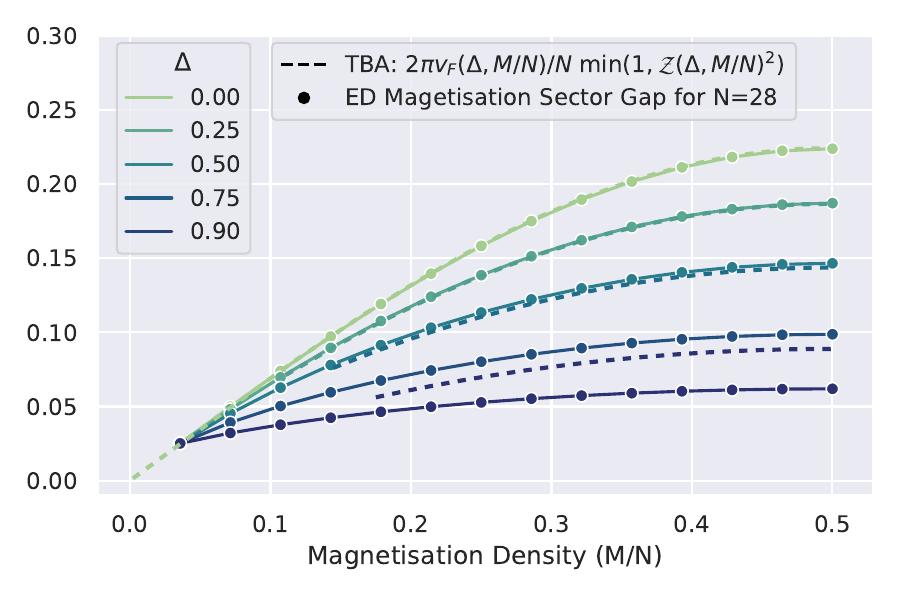}
    \caption{
    ED results for magnetization sector gap for fixed $N=28$ in comparison with iterative solutions of the TBA integral equations of $v_F, \mathcal{Z}^2$.
    The TBA large system size limit and small size ED agree (up to finite-size effects and instability of TBA numerics near $\Delta=1$~\cite{SM}), the gap closes only at $\Delta=1$ or $M/N=0$.
    As TBA numerics is limited near $\Delta = 1$, $\Delta=0.9$ is not shown and, in conjunction with finite size effects, the deviations occur~\cite{SM}.
    }
\label{fig:xxz_coefficients}
\end{figure}
The scaling relation for the lowest energy excitations in the thermodynamic Bethe ansatz contains two system-size independent values $v_F, \mathcal{Z}^2$.
For the closing of the gap, it is then relevant, how these behave with changing magnetization density $M/N$ and $\Delta$.
The thermodynamic Bethe ansatz for the XXZ model takes the form of the set of integral equations
\begin{align}
    & \mathcal{Z}(\lambda)+\frac{1}{2 \pi} \int_{-\Lambda}^{\Lambda} \mathcal{K}(\lambda-\mu) \mathcal{Z}(\mu) \mathrm{d} \mu=1
    \quad\quad
    \mathcal{K}(\lambda) =\frac{\sin (2 \gamma)}{\cosh \lambda-\cos (2 \gamma)}
    \quad\quad
    \mathcal{Z}(\Lambda)= \frac{\pi \sqrt{1-\Delta^2}}{4\gamma h \cosh \left(\frac{\pi}{2} \frac{\Lambda}{\gamma}\right)} =: \mathcal{Z}
    \label{eq:xxz_tba_integral_equation_Z}
    \\
    &\left.\begin{aligned}
    & \rho(\lambda)+\frac{1}{2 \pi} \int_{\Lambda}^{\Lambda} \mathcal{K}(\lambda-\mu) \rho(\mu) \mathrm{d} \mu 
    = \frac{1}{2\pi}\frac{d}{d\lambda} 2 \arctan \left[\cot \left(\frac{\gamma}{2}\right) \tanh \left(\frac{\lambda}{2}\right)\right]\\
    & \varepsilon(\lambda)+\frac{1}{2 \pi} \int_{-\Lambda}^{\Lambda} \mathcal{K}(\lambda-\mu) \varepsilon(\mu) \mathrm{d} \mu 
    = 2 h- \frac{\sin ^2 \gamma}{\cosh \lambda-\cos \gamma}
    \end{aligned} \quad\right\}\quad 
    \begin{aligned}
    &v_F=\left.\frac{1}{2 \pi \rho(\Lambda)} \frac{\partial \varepsilon(\lambda)}{\partial \lambda}\right|_{\lambda=\Lambda}\\
    & \int_{\Lambda}^{\Lambda} \rho(\mu) \mathrm{d} \mu = \frac{M}{N}.
    \end{aligned}
    \label{eq:xxz_tba_integral_equation_epsilon_rho}
\end{align}
As stated in the main text, an analytical solution for values of $\mathcal{Z}, v_F, M/N$ is not possible~\cite{franchini2017introduction}, so one has to resort to numerics.
Equations of the form of Eq.~\ref{eq:xxz_tba_integral_equation_epsilon_rho} can be solved by iteration method; i.e. starting with a function (e.g. a polynomial), then repeatedly numerically integrating with the kernel until convergence.
Similarly, Eq.~\ref{eq:xxz_tba_integral_equation_Z} can be approached by additionally solving the algebraic equation for the integral boundaries approximately at each step.
The entire system can then be solved by first computing $Z(\lambda), \Lambda$ by the latter, then both $\rho(\lambda), \epsilon(\lambda)$ by the former and finally integrating $\rho$ for $M/N$ and computing a numerical derivative for $v_F$.
The results of these steps are shown in Fig.~\ref{fig:tba_background}.
The dependency on $h$ can then be replaced by $M/N$, and one can take the minimum of the factors in the scaling relation to arrive at the values of $v_F, \mathcal{Z}^2$ in Fig.~\ref{fig:xxz_coefficients}.
The numerical data shows good agreement of a small-size $N=28$ system with the infinite size-limit and suggests that the coefficients do not vanish except at the phase boundary.

\section{RG Parent Hamiltonian Small $g$ Perturbation Theory}
\label{sec:rg_perturbation_theory}
For the Richardson-Gaudin model, a full proof of a lower bound for the gap of the parent Hamiltonian has been elusive, but for small $g$ (weak interactions) the behavior can be understood via perturbation theory.

To this aim, we consider Weyl's theorem \cite{weyl1912asymptotische}, which states that for two hermitian matrices $X, Y \in \mathbb{C}^{d\times d}$, $|\lambda_k(X) - \lambda_k(Y)| \leq ||X-Y||_{\mathrm{op}}$, where $\lambda_k(\cdot)$ denotes the $k$-th eigenvalue with ordering $\lambda_1 \geq \dots \geq \lambda_d$. Now, $H^{\mathrm{RG}}_\mathrm{P}(g)$ is positive-semidefinite and has energy $0$ only for the eigenstate it is constructed for, s.t. $\lambda_{2^N}(H^{\mathrm{RG}}_\mathrm{P}(g)) = 0$ and thus $\Delta = \lambda_{2^N-1}(H^{\mathrm{RG}}_\mathrm{P}(g)) > 0$ where $\Delta$ is the gap.
Moreover, from the quadratic Bethe equations $\lambda_{2^N}(H^{\mathrm{RG}}_\mathrm{P}(0)) = 0$ and $\lambda_{2^N-1}(H^{\mathrm{RG}}_\mathrm{P}(0)) = 1$.
Now we define $A^k = \frac{\sigma_k^z}{2} + \frac{1}{2}$ and $B^k = \frac{1}{4} \sum_{j \neq k}^N \frac{\vec{\sigma}_k \cdot \vec{\sigma}_j - 1}{\epsilon_k - \epsilon_j}$, such that $Q^{(k)}(g) = A^k + g B^k$.
Applying Weyl's theorem then leads to
\begin{align}
&|\Delta - 1| = |\lambda_{2^N-1}(H^{\mathrm{RG}}_\mathrm{P}(g)) - \lambda_{2^N-1}(H^{\mathrm{RG}}_\mathrm{P}(0))|
\leq ||\sum_{k=1}^{N} (A^k + g B^k - q^{(k)}_v(g))^2 - (A^k - q^{(k)}_v(0))^2||\\
&= ||\sum_{k=1}^N g^2 {B^k}^2 + g (B^k A^k + A^k B^k - 2 q_v^{(k)}(g) B^k) + q_v^{(k)}(g)^2 - q_v^{(k)}(0)^2 + 2 A^k (q_v^{(k)}(0) - q_v^{(k)}(g))||\\
&\leq \sum_{k=1}^N g^2 ||{B^k}||^2 + g (2||B^k|| + 2 |q_v^{(k)}(g)|\ ||B^k||) + |q_v^{(k)}(g) - q_v^{(k)}(0)||q_v^{(k)}(g) + q_v^{(k)}(0)| + 2 |q_v^{(k)}(0) - q_v^{(k)}(g)|
\label{eq:pt_second_last_line}\\
&\leq 4 \sum_{k=1}^N g^2 ||{B^k}||^2 + 2 g ||B^k||
\leq 4 N \left( g^2 \left(\frac{N}{\min_{k, j} |\epsilon_k - \epsilon_j|}\right)^2 + 2 g \left(\frac{N}{\min_{k, j} |\epsilon_k - \epsilon_j|}\right) \right)
\label{eq:pt_last_line}
\end{align}
where we employed submultiplicativity and the triangle inequality as well as $||A^k||=1$, $q_v^{(k)}(0) \in \{0,1\} \leq 1$, $|q_v^{(k)}(g)| \leq 1 + g ||B^k||$ and $||B^k|| \leq \frac{N}{\min_{k, j} |\epsilon_k - \epsilon_j|}$.
As $q^k_v(0)$ is $0$ or $1$ for half the eigenstates respectively,  for $g < 1/ (2 ||B^k||)$ Weyl's theorem imply that the $2^{N-1}$ smallest $q_v^{(k)}(g)$ eigenvalues (corresponding to $q^k_v(0) = 0$) are smaller than $g ||B^k||$, while the largest $2^{N-1}$ $q_v^{(k)}(g)$ eigenvalues (anagolously corresponding to $q^k_v(0) = 1$) are larger than $1 - g ||B^k||$.
By continuity of the eigenvalues \cite{kato2013perturbation} and under the assumption $g < 1/ (2 ||B^k||)$ the two groups do not overlap as $1 - g ||B^k || > g ||B^k||$.
As the ordering of the eigenvalues can thus be matched and $Q^{(k)}(g) - Q^{(k)}(0) = g B^k$, by Weyl's theorem $|q_v^{(k)}(g) - q_v^{(k)}(0)| \leq g ||B^k||$, which is employed to go to  Eq.~\eqref{eq:pt_last_line}.

Consequently, if $g < \frac{\min_{k, j} |\epsilon_k - \epsilon_j|}{2(2 + \sqrt{5}) N^2}$ (thus fulfilling the condition $g < 1/2 ||B^k||$), $|\Delta - 1| < 1$ for any $N$ which implies a $\mathcal{O}(1)$ gap of the Richardson-Gaudin Hamiltonian.

\section{RG Parent Hamiltonian Large $g$ Gap}
\label{sec:rg_gap_large_g}
We will now discuss the opposite case to the previous section and present a proof that the minimal Richardson-Gaudin parent Hamiltonian has a gap of $1/N$ in the large-$g$ (strong interactions) limit, i.e.
\begin{equation}
    \lim_{g\rightarrow \infty} \min_{v,v^\prime} ||\vec{q}_v(g) - \vec{q}_{v^\prime}(g)||_2^2 = 1/N.
\end{equation}

First, we assume $g > 0$ and rewrite $Q^{(k)} = g (B^k + \frac{1}{g} A^k)$, where we reuse the notation of the previous section.
The $B^k$ are exactly the IOMs of the integrable Gaudin magnet \cite{gaudin1976diagonalisation}, whose eigenstates we will denote by $B^k |v^\textrm{B}\rangle = q_{v^\textrm{B}}^k |v^\textrm{B}\rangle$.
For $g\rightarrow \infty$, we consider perturbation theory in $g^{-1}$, which gives $q^k_v(g) = g(q_{v^\textrm{B}}^k + g^{-1} \langle v^\textrm{B} | A^k | v^\textrm{B} \rangle + \mathcal{O}(g^{-2}))$ for the RG IOM eigenvalues.
Collecting all IOMs in a vector, we choose the notation $\vec{q}_v(g) = g \vec{b}_v + \vec{a}_v + \mathcal{O}(1/g)$.
Expanding $||\vec{q}_v(g) - \vec{q}_{w}(g)||_2^2 = || \vec{a}_v - \vec{a}_{w} + g (\vec{b}_v - \vec{b}_{w}) + \mathcal{O}(1/g)||_2^2$ we see that, denoting the first excited state of $H^\textrm{RG}_v(g)$ by $|v^\prime \rangle$, for $g\rightarrow \infty$ we can distinguish two cases: either \textit{(i)} $(\exists |w^B\rangle \neq |v^B\rangle: \vec{b}_w = \vec{b}_v) \implies \vec{b}_{v^\prime} = \vec{b}_v$ or \textit{(ii)} $(\not \exists |w^B\rangle \neq |v^B\rangle: \vec{b}_w = \vec{b}_v)$ implies that the gap of $H^\textrm{RG}_v(g)$ diverges as $g\rightarrow \infty$.
However, if $\vec{b}_{v} = \vec{b}_{v^\prime}$, the eigenvalues of $v^B$ and $v^{\prime B}$ must be identical for all $B^k$, while in the literature \cite{falqui2003gaudin,garajeu2001singular} it is known that a subset of any $N-1$ of the $B^k$ together with total magnetization $M = \sum_{i=1}^N \sigma_z$ forms a complete set of commuting observables for the Gaudin magnet - thus $v$ and $v^\prime$ must have different magnetization.
However, by Eq.~\eqref{eq:rg_gap_different_magnetization} this implies that $||\vec{q}_v(g) - \vec{q}_{v^\prime}(g)||_2^2 \geq 1/N$.
Consequently, $\min_{v,v^\prime} ||\vec{q}_v - \vec{q}_{v^\prime}||_2^2$ for some choice of $\epsilon$ and $N$ is either diverging for $g \rightarrow \infty$ in case $\not \exists |w^\mathrm{B}\rangle \neq |v^\mathrm{B}\rangle: \vec{b}_w = \vec{b}_v$ or lower bounded by $1/N$ in the other case.
However it can be seen that the Dicke states are eigenstates of all $B^k$, and thus Gaudin magnet eigenstates, which always have $\vec{b} = \vec{0}$ and $\vec{a} = \frac{M}{N} \begin{pmatrix} 1, 1, \dots, 1, 1 \end{pmatrix}^T$.
By this explicit construction, the latter case, with $\exists |v^B\rangle, |v^{\prime B}\rangle: \vec{b}_v = \vec{b}_{v^\prime}$ and the gap being lower bounded by $1/N$, is always true, which concludes the overall lower bound.
For two Dicke states with magnetizations $M$ and $M+1$ this corresponds to $||\vec{a}_v - \vec{a}_{v^\prime}||^2_2=1/N$ which is thus consistent with a $1/N$ scaling of the gap of the RG parent Hamiltonian, thus proving the bound is tight and being consistent with the main text numerics.
Moreover these values of $\vec{a}$ and $\vec{b}$ correspond exactly to the minimal pairs of vectors found in the numerics discussed in Sec.~\ref{sec:rg_numerics_details}.

\section{RG Numerics Details}
\label{sec:rg_numerics_details}

\begin{figure}
\subfloat[
\label{fig:delta_g_system_size_scaling}
Numerical scaling of minimal $\Delta g$ for going to $g = 10$ with the dynamical step adaptation described in Sec.~\ref{sec:rg_numerics_details}, without using Taylor expansion or by expanding to first order.
While not conclusive, for the former this suggests $\Delta g = \mathcal{O}(1/2^N)$ while for the latter this is alleviated.
]{\includegraphics[height=5.5cm]{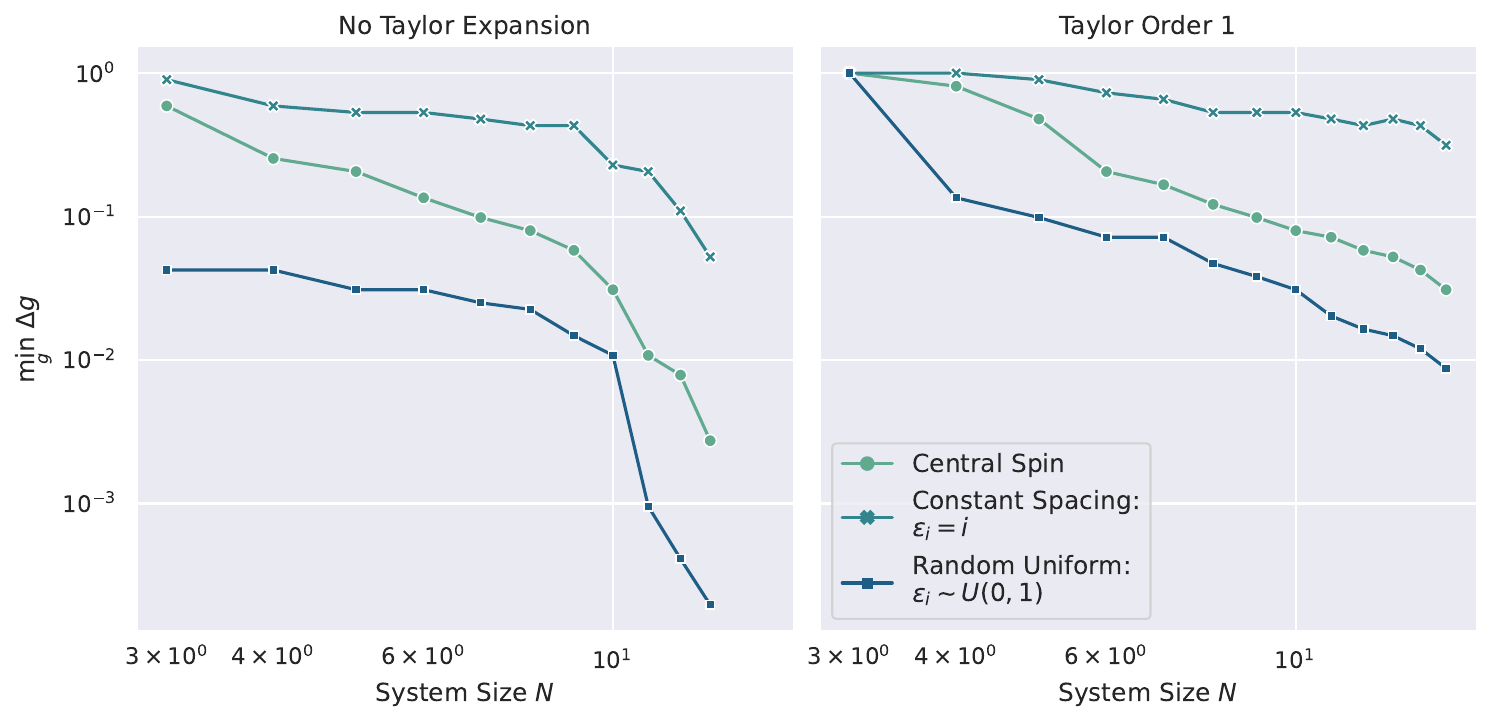}}
\hspace{1em}
\subfloat[
\label{fig:taylor_condition_number_scaling}
Maximum of the condition number of the linear system determining the first derivative of the quadratic Bethe equations over all eigenvectors and the entire adiabatic path.
]{\includegraphics[height=5.5cm]{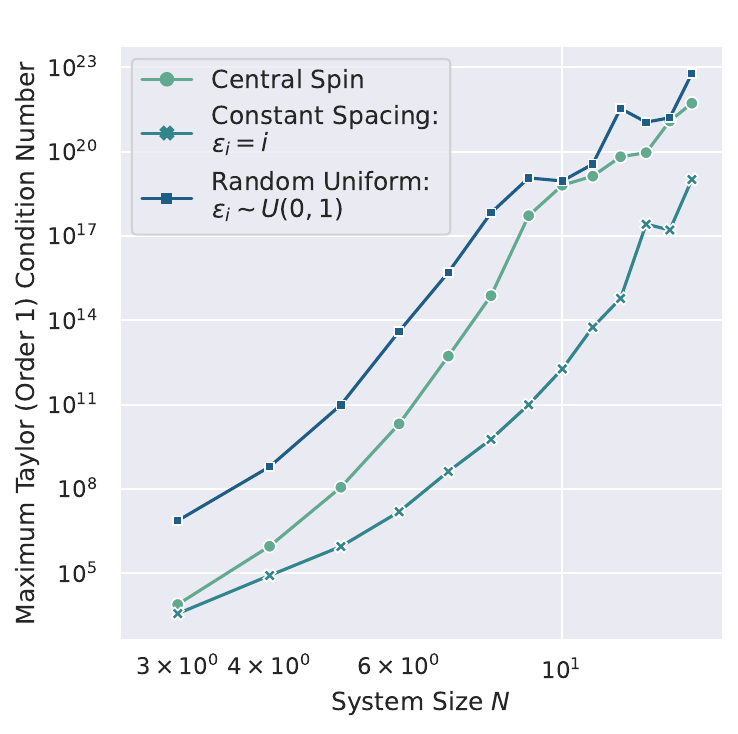}}\\

\subfloat[
\label{fig:delta_g_g_scaling}
Same data as Fig.~\ref{fig:delta_g_system_size_scaling} for Taylor expansion of first order, but shown against $g$.
$\Delta g$ increases linearly for growing $g$.
]{\includegraphics[height=5.5cm]{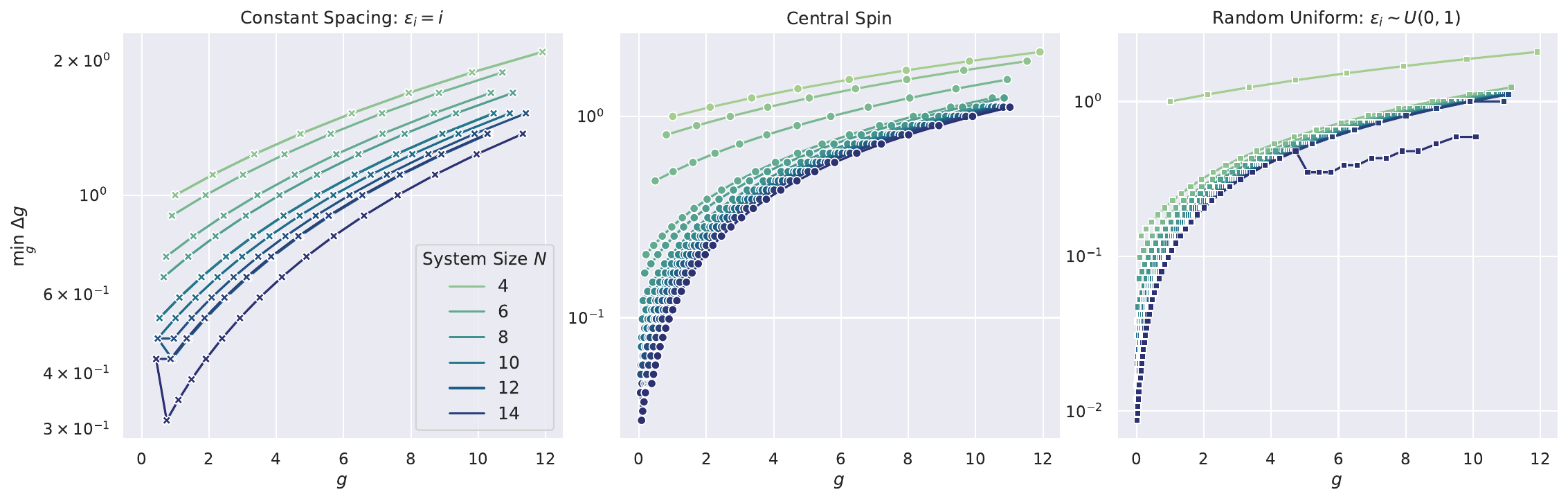}}
\caption{Numerical behavior of quantities relevant to classical complexity of Richardson-Gaudin eigenvalue method: scaling of iteration step size $\Delta g$ and condition number in case of Taylor expansion; for details see Sec.~\ref{sec:rg_numerics_details}.
}

\end{figure}

\begin{figure}
    \centering
    \includegraphics[width=\linewidth]{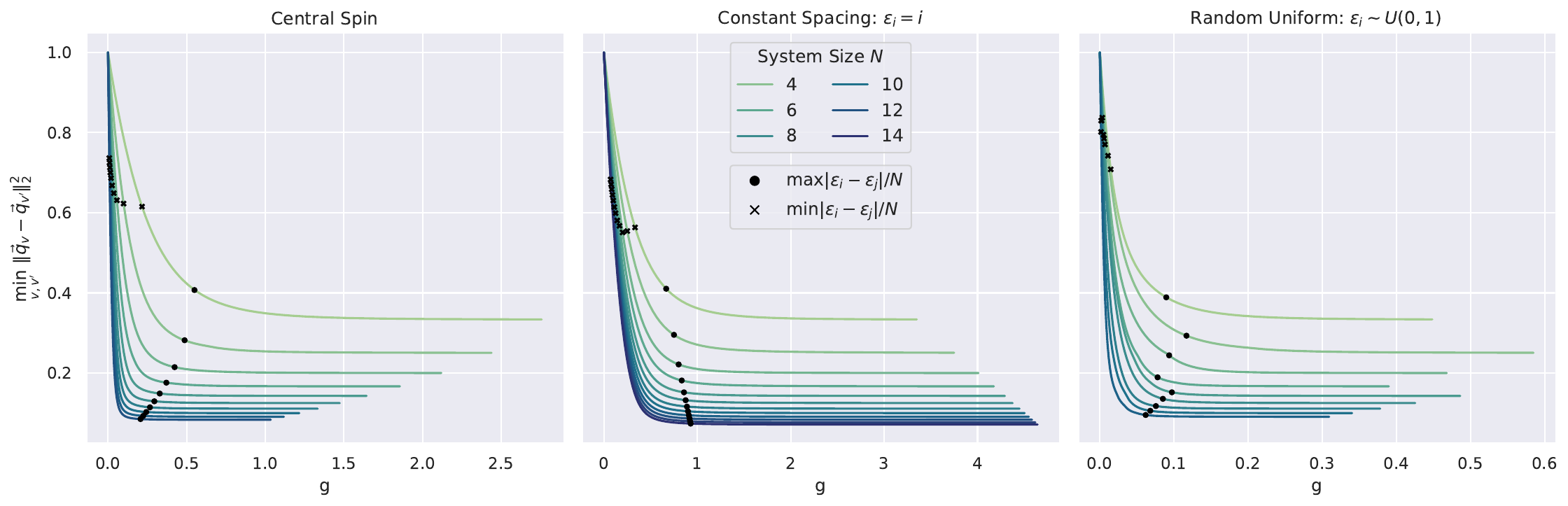}
    \caption{$H_v^{\mathrm{RG}}$ smallest gap minimized over all $v$ (not minimized over the adiabatic path) as a function of $g$. Black markers show relevant orders for norms of interacting and non-interacting part of $Q^{(k)}$; see Sec.~\ref{sec:rg_numerics_details} for further details.}
    \label{fig:min_distance_g}
\end{figure}

\begin{figure}
    \includegraphics[width=\linewidth]{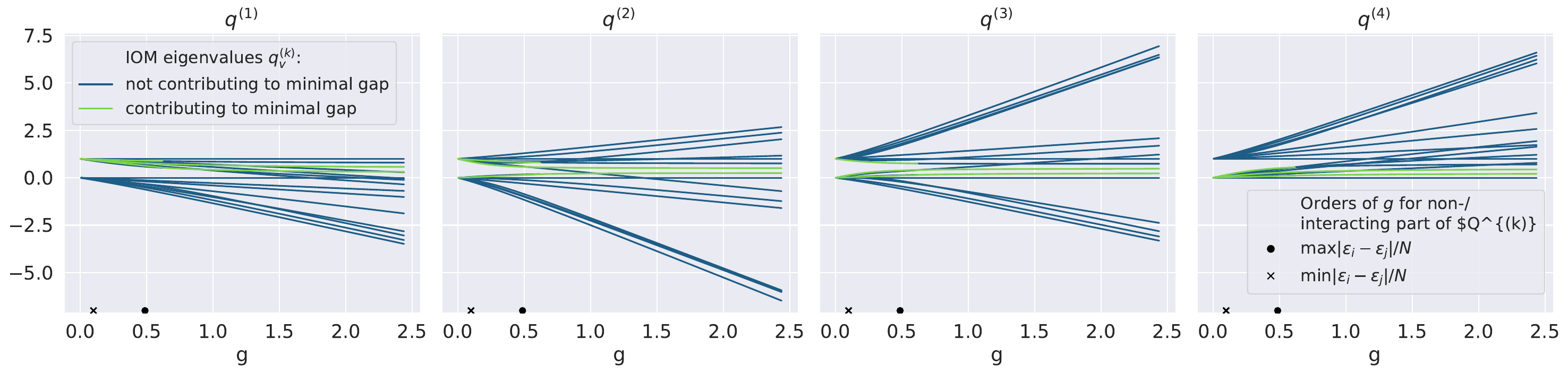}
    \caption{IOM eigenvalues $\vec{q}_v(g)$ shown against $g$ as an explicit example for small system size $N=4$.
    Each line corresponds to one eigenstate.
    Marked color corresponds to all eigenvectors, whose pairwise distances correspond to the overall minimum of the parent Hamiltonian gap at the value of $g$.
    }
    \label{fig:rg_eigenvalues_g}
\end{figure}

\begin{figure}
    \centering
    \includegraphics[width=.8\linewidth]{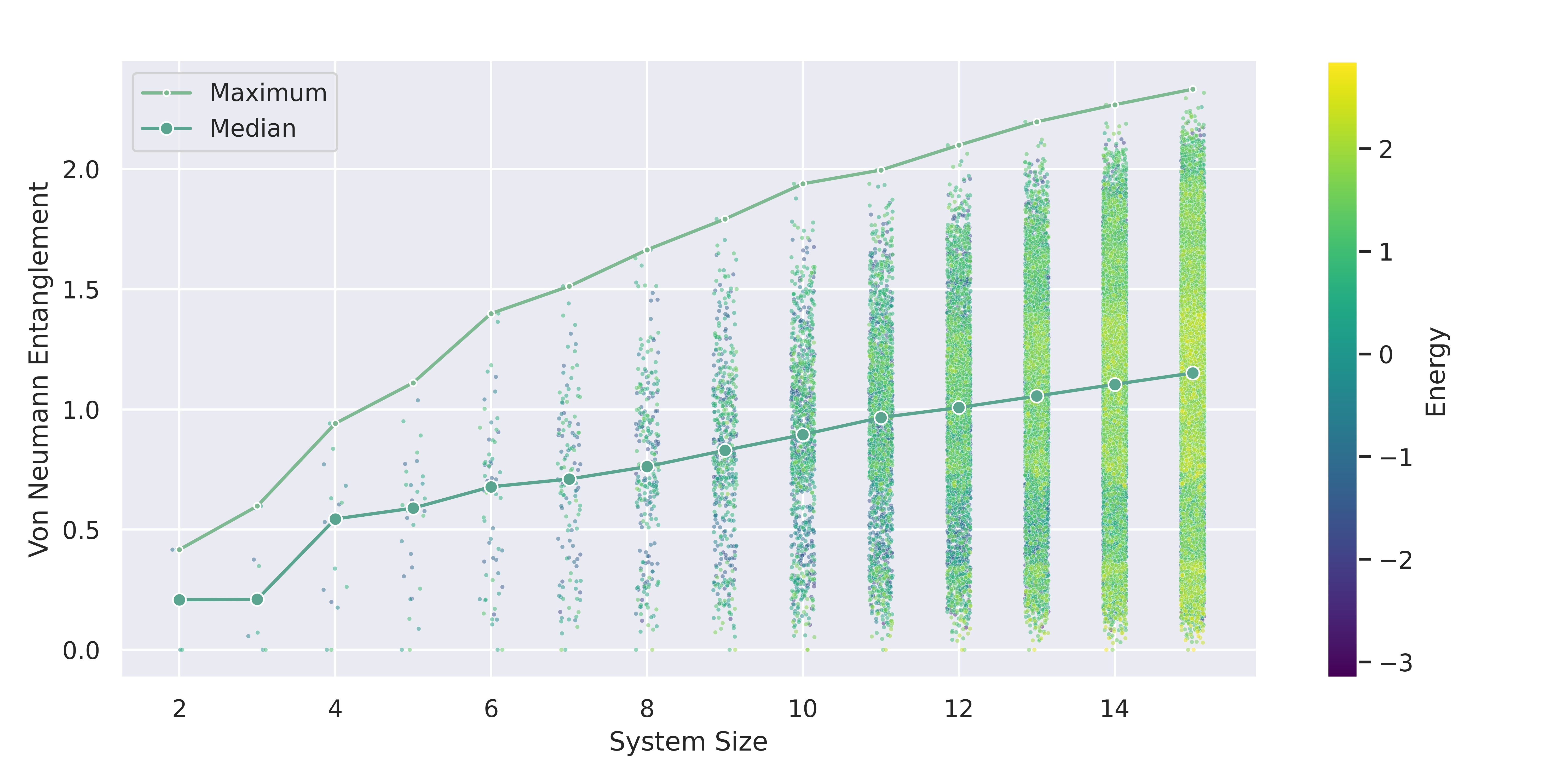}
    \caption{
Numerical results for bipartite von Neumann entanglement entropy of central-spin model for all eigenstates against system size.
The bipartition is between the $N/2$ more strongly and $N/2$ less strongly interacting sites; the central spin model is taken at $g = 1$.
Our numerical results do not allow us to support either a $O(\log(N))$ of $O(N)$ scaling .
    }
    \label{fig:central_spin_entanglement_scaling}
\end{figure}

The numerical method for the computation of all IOM eigenvalues corresponding to an RG eigenstate that was mentioned in the main text was implemented using a Levenberg–Marquardt algorithm as the local optimizer for the central spin, constant spacing and random uniform choices of the $\epsilon_i$.
As we are interested in the minimal gap of $H^\mathrm{RG}_\mathrm{v}$ for all eigenstates $v$, the computation needs to be performed for all binary vectors as starting values.
To simultaneously find the maximal $\Delta g$ guaranteeing that the local solver does not leave the local minimum at any iteration step, we adapt $\Delta g$ dynamically in the following way: \textit{(i)} after any iteration, the pairwise distance between all IOM eigenvalue vectors $\vec{q}_v$ and the error resulting from the local optimizer are computed. \textit{(ii)} If the former is too small (\textit{i.e.} one eigenstate local minimum was left and the solver found the same solution twice), or the latter is too large (\textit{i.e.} the optimizer converged to a local minimum, which does not correspond to a true solution of the quadratic Bethe equations) $\Delta g$ is multiplied with an adaption factor $<1$. \textit{(iii)} If the step succeeds, we attempt to increase $\Delta g$, by dividing by the adaption factor.
Thus $\Delta g$ is adapted exponentially quickly to correspond to the largest possible value of $\Delta g$, at which the convergence radii of the local optimizer around all solutions of the algebraic equations are never left.
As an additional optimization, the initial vector of the local optimizer can instead be chosen by computing the Taylor expansion at the previous vector to a fixed order.
As for the quadratic Bethe equations, the required derivatives reduce to linear sets of equations \cite{PhysRevB.83.235124}.
Specifically, we choose an adaption factor of $0.9$ and a maximal tolerance for the quadratic error in the quadratic Bethe equations of $10^{-14}$, while we Taylor expand to first order.
To ensure the gap is sampled at every point of the adiabatic path - as required for our proposed algorithm - we additionally set a fixed upper threshold for the step size $\Delta g$ in some circumstances, as further explained below.
This procedure results in the $\mathcal{O}(1/N)$ gap scaling data discussed in the main text Fig.~\ref{fig:rg_minimal_gap_scaling} when minimizing over both the entire adiabatic path as well as all pairs of eigenvectors.
We will now discuss the numerical results in more detail and beyond both these minimizations.

Fig.~\ref{fig:delta_g_system_size_scaling} shows how the $\Delta g$ step resulting from the dynamical adaptation described above changes with system size, when we take its minimum over the adiabatic path, both for the algorithm without Taylor expansion and with Taylor expansion to first order.
While the numerics are limited by system size, it can be seen that $\Delta g$ in general does become small and even suggests a $\Delta g = \mathcal{O}(1/\exp{(N)})$ scaling for the case without Taylor expansion.
This appears to be alleviated in the case of Taylor expansion to first order.
While the required first derivative can be computed by inverting a linear system in this case, Fig.~\ref{fig:taylor_condition_number_scaling} suggests, however, that the condition number (defined as the ratio of the largest and the smallest singular value) of the respective linear system increases with $N$, which presents a potential challenge for this approach.
Fig.~\ref{fig:delta_g_g_scaling} shows that $\Delta g$ increases with $g$, thus being consistent with the $\mathcal{O}(\log(g))$ scaling reported in the literature \cite{fecteau2022near} and suggesting large $g$ are no obstacle for the method.
Overall, our numerical results do not allow us to conclude whether our numerical method to solve for the IOM eigenvalues of the Richardson-Gaudin models is classically efficient; to our knowledge no formal complexity result is known, while the method has been applied very successfully in the literature.
It is an interesting question for future research to study this behavior analytically and investigate whether this complexity is optimal.
The extensive literature on homotopy methods for the solution of algebraic equations as well as certified bounds for Newton-type optimizers (c.f. Sec.~\ref{sec:smale_alpha_theory}) might be relevant in this regard.

Returning to the scaling of the parent Hamiltonian gap, Fig.~\ref{fig:min_distance_g} shows the minimal gap of $H_v^\mathrm{RG}$ over all eigenstates, plotted against $g$.
It can be seen that the gap decreases monotonically for all three models and converges to a fixed value.
Indeed, inspecting the integrals of motions of Eq.~\eqref{eq:rg_Qk} and upper/lower bounding the norms of the interacting and non-interacting parts, the two parts become of comparable order at the earliest for $g = \min_{i,j} | \epsilon_i - \epsilon_j| / N$ and for $g = \max_{i,j} | \epsilon_i - \epsilon_j| / N$ at the latest.
The final gap value then apparently corresponds to where the interacting part dominates.
This observation gives reason to sample the entire adiabatic path by setting the maximal $\Delta g$ threshold to a fraction of $\min_{i,j} | \epsilon_i - \epsilon_j| / N$ and to simulate to a $g$ to some multiple of $\max_{i,j} | \epsilon_i - \epsilon_j| / N$, as this guarantees not missing any behavior in the crossover region of $g$ and including the strongly interacting regime.
Indeed, for all system sizes $N \leq 11$ in Fig.~\ref{fig:rg_minimal_gap_scaling} (and all curves in Fig.~\ref{fig:min_distance_g}) the maximal step size threshold is set to $\min_{i,j} | \epsilon_i - \epsilon_j| / 20 N$ and the simulation run to $5 \max_{i,j} | \epsilon_i - \epsilon_j| / N$.
For $N > 11$ in Fig.~\ref{fig:rg_minimal_gap_scaling}, no upper $\Delta g$ threshold is set for reasons of computational cost, however the monotonic behavior in Fig.~\ref{fig:rg_minimal_gap_scaling} strongly suggests that this does not limit the relevance for arbitrary $g$ values.

Fig.~\ref{fig:rg_eigenvalues_g} now additionally goes beyond the minimization over pairs of eigenvectors, and shows the IOM eigenvalues changing with $g$ for all eigenvectors for a selected small system size of $4$.
After the values start from the binary vectors at $g=0$, the eigenvalue vectors enter a regime with crossings between the $q^{(k)}(0)=0/1$ eigenvalues.
A physical intuition for the resulting parent Hamiltonian gap in this picture might be the emergence of an avoided crossing.

A further important question is whether the RG models have volume law states anywhere in their spectra.
As a numerical evaluation of this question, Fig.~\ref{fig:central_spin_entanglement_scaling} shows an exact diagonalization finite-size scaling of the von Neumann entropy for the central spin model. The necessarily small system sizes make the results inconclusive between a $O(\log(N))$ and $O(N)$ entanglement scaling.

\section{Smale's $\alpha$-theory}
\label{sec:smale_alpha_theory}
\begin{figure}
    \centering
    \includegraphics[width=\linewidth]{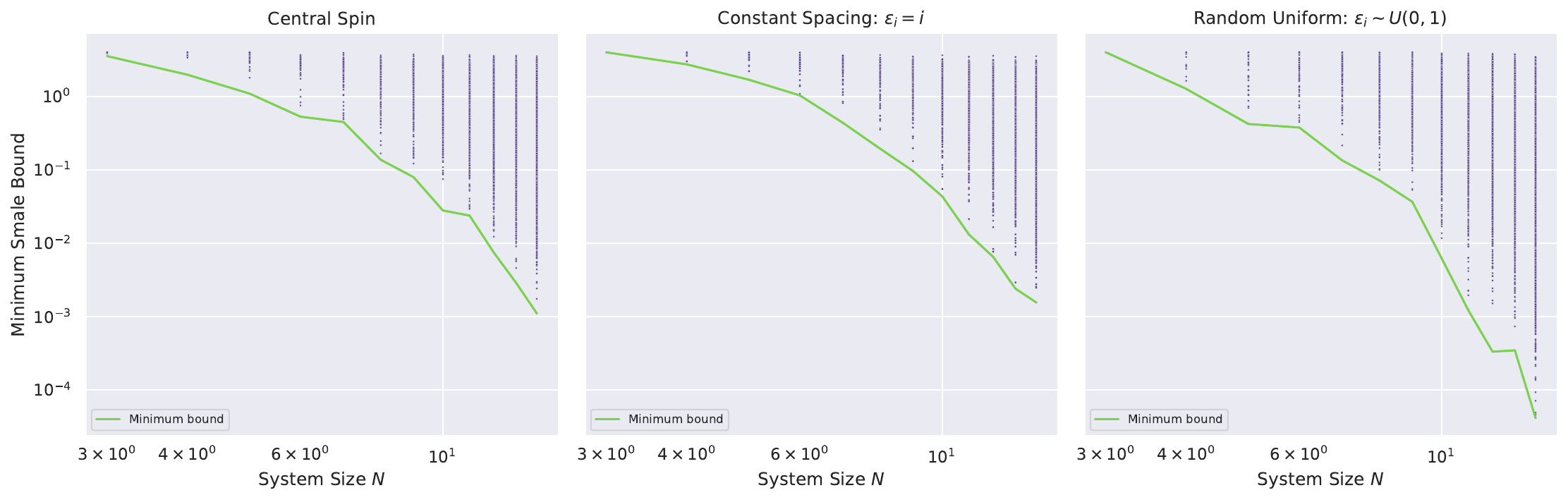}
    \caption{Lower bound for parent Hamiltonian gap via Smale bound.
    Dots corresponds to values of Eq.~\ref{eq:smale_rg_lower_bound} for all eigenstates of the model.
    It appears that the minimum over all eigenstates decreases exponentially.
    }
    \label{fig:smale}
\end{figure}
With the quadratic Bethe equations available for the RG model, the question of the gap of the Hamiltonian becomes a question of the minimal geometric distance between any root vectors of this set of algebraic equations, as $H^\mathrm{RG}_\mathrm{P}|v^\prime\rangle = ||\vec{q}_v-\vec{q}_{v^\prime}||_2^2$.
This question has been considered in the mathematical literature, and different lower bounds are known.
Importantly, the so-called Smale's $\alpha$-theory~\cite[Chs. 8-14]{blum1998complexity}, originally developed to give bounds on the convergence radius of Newtons algorithm around the root of a function, appears to be relevant.
In fact, for an analytic function $f: E \rightarrow F$ mapping between Banach spaces $E, F$, for any two roots $f(\zeta) = f(\zeta^\prime) = 0$, $\zeta \neq \zeta^\prime$, one has the following Smale's bound
\begin{equation}
    \label{eq:smale_bound}
    ||\zeta - \zeta^\prime|| \geq \frac{5-\sqrt{17}}{4} \frac{1}{\sup_{k\geq 2} ||\frac{D f(\zeta)^{-1} D^k f(\zeta)}{k!}||^{1/(k-1)}}
\end{equation}
if the inverse of the Jacobi matrix $Df(\zeta)$ exists.
For the class of RG models that were considered in the main text, the equations are all polynomials of degree $2$.
For these equations the supremum over $k$ in Eq.~\ref{eq:smale_bound} drops out, as the $D^k$ on such polynomials becomes zero for $k > 2$ and the tensors can be computed explicitly:
\begin{align}
    &(Df(\zeta))_{k,l} = \partial_k f_l(\zeta) = \begin{cases}
        i \leq N: &
        \begin{cases}
            k = l: & 2q^{(k)} - 1 + \frac{g}{2} \sum_{k\neq j}^N \frac{1}{\epsilon_k - \epsilon_j}\\
            k \neq l: & -\frac{g}{2}\frac{1}{\epsilon_l - \epsilon_k}
        \end{cases}\\
        i = N+1: & 1
    \end{cases}\\
    &(D^2f(\zeta))_{k,l,m} = \partial_k \partial_l f_m(\zeta) = \begin{cases}
        k = l = m \leq N: & 2\\
        \text{else:} & 0
    \end{cases}.
\end{align}
Lower bounding the Smale bound via submultiplicativity and using that $||M^{-1}|| = 1/\sigma_{min}(M)$, one arrives at
\begin{equation}
    ||\zeta - \zeta^\prime|| \geq \frac{5-\sqrt{17}}{2}\frac{\sigma_{min}(D f(\zeta))}{||D^2 f(\zeta)||},
    \label{eq:smale_rg_lower_bound}
\end{equation}
where $||D^2 f(\zeta)|| = 2$.

At this point we note the similarity to the Gaudin determinant expression \cite{Gaudin,1982.Korepin.CMP.86,claeys2017inner} and the connection to the condition number of the linear system occurring in the Taylor expansion for the eigenvalue based solution of the quadratic Bethe equations discussed in Sec.~\ref{sec:rg_numerics_details}.

However, even as the resulting tensors have a lot of structure, as the Smale bound is still dependent on the solution vector of the quadratic Bethe equations, an analytical lower bound is not immediate.
A numerical analysis can be done by calculating the Smale bounds for all eigenvectors for small system sizes, by using the explicit results of the numerical method described in the previous section.
Fig.~\ref{fig:smale} shows these results for the central spin model.
Unfortunately, the minimal Smale bound appears to be exponentially small in system size and thus insufficient for a proof of the gap scaling observed in the numerics.
Yet, the numerics suggests the Smale bound does not decay for many eigenstates, and it is thus an interesting question for future research whether methods of this sort could be used to prove efficient preparation for some eigenstates of the RG models, or if the Smale bound is tighter for other classes of integrable models.

\end{document}